\documentclass[final,3p,times,preprint]{elsarticle}

\usepackage{amssymb}

\usepackage{graphicx,epsfig}
\usepackage{upgreek}
\usepackage[version=3]{mhchem}
\usepackage{lmodern} 
\usepackage[T1]{fontenc} 
\usepackage{multirow}

\usepackage{tabularx}
\newcolumntype{C}[1]{>{\centering\arraybackslash}p{#1}}

\newcommand{\mk}[1]{\,$^{\rm #1)}$}
\newcommand{\degr}{$^{\circ}$}

\newcommand{\etalp}{\textit{et~al.~}}

\usepackage{color}
\newcommand{\add}[1]{\textcolor{black}{#1}}
\newcommand{\rem}[1]{\textcolor{blue}{}}

\journal{Colloids and Surfaces A}

\begin{document}

\begin{frontmatter}



\title{Capillary rise dynamics of liquid hydrocarbons in mesoporous silica as explored by gravimetry, optical and neutron imaging: Nano-rheology and determination of pore size distributions from the shape of imbibition fronts}
\author{Simon Gruener}
\address{Experimental Physics, Saarland University, D-66041 Saarbruecken (Germany)}
\address{Sorption and Permeation Laboratory, BASF SE, D-67056 Ludwigshafen (Germany)} 
\author{Helen E. Hermes}
\address{Condensed Matter Physics Laboratory, Heinrich Heine University, D-40225 Duesseldorf (Germany)}

\author{Burkhard Schillinger} 
\address{FRM II and Physik-Department E21, Technische Universitaet Muenchen, James-Franck-Str., 85748 Garching (Germany)}

\author{Stefan U. Egelhaaf}
\address{Condensed Matter Physics Laboratory, Heinrich Heine University, D-40225 Duesseldorf (Germany)}

\author{Patrick Huber}
\address{Experimental Physics, Saarland University, D-66041 Saarbruecken (Germany)}
\address{Institute of Materials Physics and Technology, Hamburg University of Technology, D-21073 Hamburg-Harburg (Germany)}

\begin{abstract}
We present combined gravimetrical, optical, and neutron imaging measurements of the capillarity-driven infiltration of mesoporous silica glass (Vycor) by hydrocarbons. Square-root-of-time Lucas-Washburn invasion kinetics are found for\rem{selected} linear alkanes from n-decane (C10) to n-hexacontane (C60) and for squalane, a branched alkane, in porous monoliths with 6.5 nm or 10 nm mean pore diameter, respectively. Humidity-dependent experiments allow us to study the influence on the imbibition kinetics of water layers adsorbed on the pore walls. Except for the longest molecule studied, C60, the invasion kinetics can be described by bulk fluidity and bulk capillarity, provided we assume a sticking, pore-wall adsorbed boundary layer, i.e. a monolayer of water covered by a monolayer of flat-laying hydrocarbons. For C60, however, an enhanced imbibition speed compared to the value expected in the bulk is found. This suggests the onset of velocity slippage at the silica walls or a reduced shear viscosity\rem{as a function of all-trans length of the hydrocarbon backbone and thus a} \add{due to the} transition towards a behaviour typical of polymer-like flow in confined geometries.  Both\add{,} light scattering and neutron imaging\add{, indicate}\rem{evidence} a pronounced roughening of the imbibition fronts. Their overall shape and increase in width can be resolved by neutron imaging. The fronts can be described by a superposition of independent wetting fronts moving with pore size-dependent square-root-of-time laws and weighted according to the pore size distributions obtained from nitrogen gas sorption isotherms. This finding indicates that the \add{shape of the} imbibition front\rem{shape} in a porous medium, such as Vycor glass, with interconnected, elongated pores, is solely determined by independent movements of liquid menisci. These are dictated by the Young-Laplace pressure and hydraulic permeability variations and thus the pore size variation at the invasion front.\rem{By the same token} Our results suggest that\rem{it is possible to derive} pore size distributions \add{can be derived} from the broadening characteristics of imbibition fronts. 

\end{abstract}


\end{frontmatter}

\section{Introduction}
Confinement of liquids in pores plays a dominant role in phenomena ranging from clay swelling, oil recovery, wet adhesion and catalysis, to drug delivery, self-assembling, protein folding and transport across artificial nanostructures, bio-membranes and tissues \cite{Drake1990, Lenormand1990, Christenson2001, Clarke2002, Alba-Simionesco2006, Knorr2008, Binder2008, Perkin2013, Berg2013, Kriel2014, Datta2014, Jiang2014, Murison2014, Bon2014, Xue2015, Zhang2015, Stroock2014,Cherstvy2015, Ichilmann2015, Huber2015}. Porous media are also gaining an increasing relevance in template-assisted (electro-)deposition of nano structures \cite{Huczko2000, Yin2001, Steinhart2002, Sander2003} and in the synthesis of soft-hard hybrid materials \cite{Coakley2003, Ford2005, Hoffmann2006, Thomas2008, Kityk2010,Sousa2014, Martin2014} by melt-infiltration \cite{Jongh2013}, for example in the field of battery and supercapacitor design \cite{Westover2014} and for the preparation of multifunctional structural materials \cite{Elbert2014, Wang2013}. 

In particular, transport in nanoporous media is of relevance in micro- and nanofluidics \cite{Eijkel2005, Hoeltzel2007, Gruener2008, Squires2005, Urbakh2004, Stone2004, Majumder2005,  Dittrich2006, Whitby2007, Persson2007, Schoch2008,  Piruska2010, Kirby2010, Koester2012, Bocquet2014, Vincent2014, Li2015a}. It is both of scientific and technological interest, whether macroscopic wetting and electrowetting properties \cite{Xue2014, Xue2015a} or values of fluid parameters, such as the viscosity $\eta$, surface and interfacial tensions $\sigma$, accurately describe a liquid down to very small length scales, on the order of the size of its building blocks \cite{Fradin2000, Vinogradova2011, Seveno2013, Vincent2015}. Measurements with the surface-force apparatus (SFA)\rem{, which} allow one to study shear viscosities \cite{Chan1985, Christenson1982, Stevens1997, Georges1993, Heinbuch1989, Ruths2000, Raviv2001} and frictional properties \cite{Rosenhek-Goldian2015,Jee2015,Israelachvili2015} of thin films with thicknesses down to sub-nanometers\add{. They} have revealed that in confinement, depending on the shear rate, the type of molecule and the surface chemistry\add{,} viscosity values quite different from the bulk values or in remarkable agreement with the bulk ones can be found. 

Moreover, given for example the small number of molecules in a crosssection of a nano pore \add{and the importance of wall effects}, the validity of the continuum approach of classical hydrodynamics is questionable in restricted geometries.\rem{Also} The velocity profile in the proximity of the confining walls plays a crucial role in the determination of the overall transport rates in nanoporous media. Today, it is established that the core concept of ``no-slip at the wall'' is valid only\rem{,} provided that certain conditions are met: a single-component fluid, a wetted surface, and low levels of shear stress. In many engineering applications these conditions are not fulfilled and studies sensitive to the near-\add{wall}\rem{surface} velocity profiles have revealed that slippage, that is a finite velocity of the liquid at the wall can occur in systems with surfactants, at high shear rates, for low wettability and low roughnesses of the confining walls \cite{Pit2000, Cheikh2003, Schmatko2005, Neto2005, Muller2008, Servantie2008, Sendner2009, Baeumchen2012, Bocquet2014}.


Pioneering experiments to probe transport behaviour through mesoporous media were performed by Nordberg \cite{Nordberg1944} and Debye and Cleland in the mid of the last century \cite{Debye1959}. Nordberg studied water and acetone flow, whereas Debye and Cleland reported on the flow of a series of linear hydrocarbons (n-pentane to n-octadecane) through mesoporous silica (Vycor glass, mean pore radius $r_0=3.5$~nm). Flow rates in agreement with Darcy's law, the generalisation of Hagen-Poiseuille's law for simple capillaries towards porous media \cite{Gruener2009, Gruener2011}, were observed. As Abeles \etalp \cite{Abeles1991} demonstrated by an experimental study on toluene using also nanoporous Vycor glass, depending on the size of the pores and on the temperature and pressure of the fluid, flow in porous media can be through gas (or Knudsen diffusion \cite{Gruener2008}), surface diffusion, and viscous liquid flow driven by capillary forces (termed ``spontaneous imbibition'') or by external hydraulic pressure (called ``forced imbibition''). 

In the following, we focus on spontaneous imbibition of selected linear hydrocarbons (normal alkanes) and one branched alkane (squalane) in a monolithic mesoporous medium. We\rem{document} \add{demonstrate} that by a combination of gravimetrical, optical and neutron imaging experiments detailed insights with regard to the capillarity-driven flow properties of hydrocarbons in this mesoporous glass can be gained. Moreover, we present evidence that the roughening of the imbibition front in a network of elongated pores with random radii, such as Vycor, contains important information on the pore size distribution of the porous medium. In fact, based on the observed imbibition kinetics, we derive a pore size distribution from the broadening kinetics which is in remarkable agreement with the pore size distribution derived from nitrogen gas sorption experiments.




\section{Theory of capillary rise dynamics}
\begin{figure} \center
\includegraphics[width=.8\columnwidth]{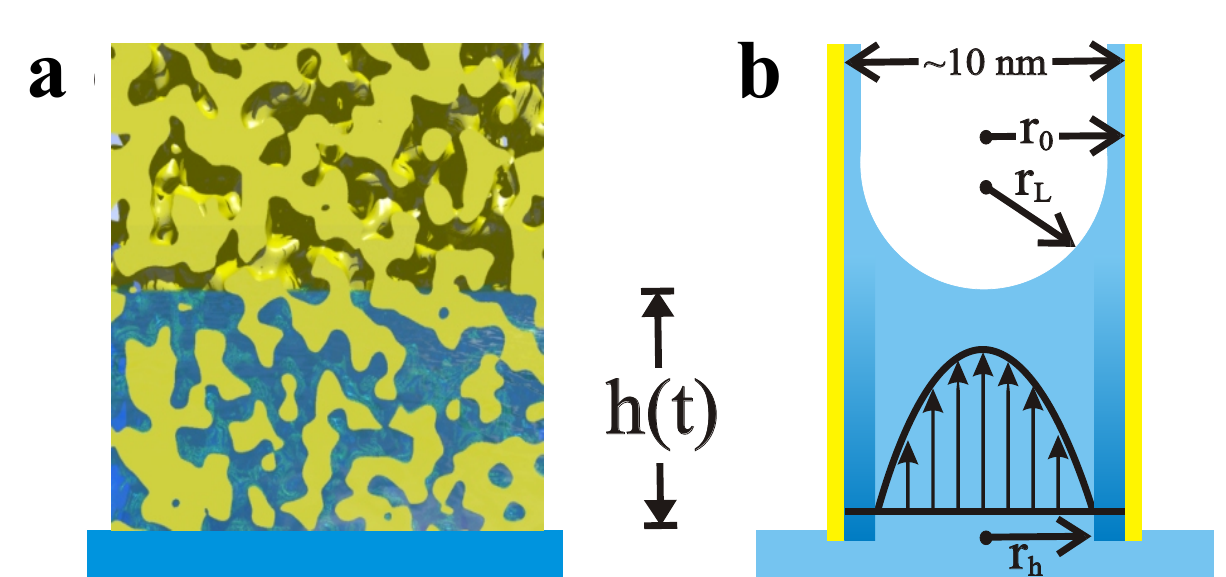}
\caption{(color online). (a)\rem{Raytracing} Illustration of spontaneous imbibition in Vycor\rem{in comparison with} \add{and} (b) schematic\rem{side view on the} \add{cross-section of an individual capillary illustrating} capillary rise of a liquid \add{in a capillary with a preadsorbed water layer}. A liquid column has advanced up to the height $h(t)$ and a parabolic\rem{fluid} velocity profile\rem{along with preadsorbed water layers beyond $h(t)$ and} \add{developed while} a\rem{shaded resting} boundary layer \add{(shaded region) remains at rest}\rem{are sketched for the nanocapillary in panel (a)}. The different radii are discussed in the text.
}
\label{fig1}
\end{figure}

The capillary rise of a wetting liquid beyond its bulk reservoir is a well-known phenomenon. From the physicist's point of view it is\rem{an impressive} \add{a classical} example of interfacial physics. Its driving force is given by the Laplace pressure 
\begin{equation}
p_{\rm L}=\frac{2\,\sigma\,\cos\theta}{r_{\rm L}} 
\label{eq:Laplace}
\end{equation}
acting on the curved meniscus of the liquid in a pore with radius $r_0 \ge r_{\rm L}$ \add{and contact angle $\theta$}. The\rem{reduction of} effective \add{radius} $ r_{\rm L}$ \add{is reduced} with respect to $ r_{\rm 0}$\rem{is induced by} \add{because of} the adsorption of water molecules on the pore walls prior to contact start -- see Fig.~\ref{fig1}\add{(b)}\rem{(a)} for a schematic illustration. For\rem{the handling of} high-energy surfaces like silica under standard laboratory conditions (humidity between 20~\% and 40~\%) such condensation processes are inevitable and not negligible in practice. Therefore, the\rem{explicit differentiation} \add{difference} between $r_0$ and $ r_{\rm L}$ is important\rem{at this point}. Of course, this effect also causes a reduction of the available volume porosity ($\phi_0 \to \phi_{\rm i}$ with $\phi_0 \ge \phi_{\rm i}$)\rem{because of the loss of with water preallocated} \add{since} pore space \add{is occupied by water and hence no longer available}. The exact value of the\rem{so-called} initial porosity $\phi_{\rm i}$ is\rem{afterwards} accessible by a thorough analysis of the overall mass uptake of the\rem{respective} sample.

For a typical alkane with surface tension $\sigma \approx 30$~mN/m (and $\cos\theta = 1$) Eq.~\eqref{eq:Laplace} yields pressures up to 200~bar in pores with radii in the nanometer range. As a direct consequence\rem{gravitation and, hence,}, the hydrostatic pressure\rem{in} \add{of} the order of some millibars (for\rem{the here regarded} the rise heights of a few centimeter\add{s considered here}) can be neglected in the description of the imbibition process \add{so that gravity is not important}. Hence, the driving pressure $\Delta p$ is solely determined by the Laplace pressure: $\Delta p = p_{\rm L}$.

\rem{At} This\rem{, the reader should bear in mind that the latter aspect is contrary} \add{is in strong contrast} to common capillary rise experiments with channel diameters in the order of some hundred micron. Here, after \add{the liquid has advanced} some centimeter\add{s,}\rem{rise height} the hydrostatic pressure compensates the Laplace pressure and the rise process comes to a standstill. Moreover, the rise dynamics are so fast that -- without any additional instruments --\rem{commonly} one\rem{may} can only observe the static equilibrium configuration at the end of the process. Experiments with mesoporous Vycor, by contrast,\rem{theoretically} yield rise heights of some kilometers \cite{Caupin2008}. Furthermore,\rem{the} \add{due to the} much smaller pores\add{,}\rem{induce a lot more} viscous drag \add{is much more important and slows down}\rem{that reduces} the overall rise \add{kinetics}\rem{dynamics} significantly. Typical rise times in our experiments are of the order of some hours or even days. This fact allows for\rem{an easy} \add{a straight-forward} recording of the dynamics by different means and a\rem{thorough subsequent} \add{detailed quantitative} analysis.

In order to describe the kinetics of the imbibition process one needs to consider the flow behaviour of a liquid in a porous substrate like Vycor. According to Ref.~\cite{Debye59} this can be done applying Darcy's law which states that the volume flow rate $\dot V = \frac{{\rm d}V}{{\rm d}t}$ normalized by the sample's cross-sectional area $A$ is given by
\begin{equation}
\frac{1}{A} \, \dot{V} = \frac{\phi_0 \, r_{\rm h}^4}{8\,r_0^2\,\tau\,\eta\, d} \Delta p \; ,
\label{eq:Darcyslaw}
\end{equation}
\rem{Here,} \add{where} $d$ is the sample's height along which the driving pressure $\Delta p$ is applied\add{, $\eta$ the viscosity, $\tau$ the tortuosity}, and $r_{\rm h}$   the hydrodynamic radius of the pore. The latter coincides with the radius over which the flow profile is established in the pore -- see Fig.~\ref{fig1}\add{(b)}\rem{(a)} for a schematic illustration. It does not necessarily have to agree with the pore radius $r_0$ because of either strongly adsorbed, immobile boundary layers of thickness $\Delta$ ($r_{\rm h}<r_0$, i.e. $r_h=r_0-\Delta$), or due to velocity slippage at the pore walls ($r_{\rm h}>r_0$). Only for the standard no-slip boundary condition is $r_{\rm h} \equiv r_0$.

It is instructive to consider the network being composed of $n$ cylindrical (isotropically oriented) pores with radius $r_0= r_{\rm h}$,\rem{so} \add{i.e.} neglecting the size distribution of the \add{pores}\rem{channels} and assuming a no-slip boundary condition. The sample's volume porosity can then be expressed by $\phi_0 =\frac{n\,\pi\,r_0^2}{A}$ and Eq.~\eqref{eq:Darcyslaw} \add{becomes:}\rem{transforms as follows}
\begin{equation}
\dot{V} = \frac{n}{\tau} \; \underbrace{\frac{\pi\,r_0^4}{8\,\eta\,d} \, \Delta p}_{\dot{V}_{\rm HP}} \; .
\label{eq:DarcyHP}
\end{equation}

The latter relation\rem{eminently} illustrates the link between Darcy's law (for a porous substrate with tortuo\add{u}s pores) and the single-channel flow rate $\dot{V}_{\rm HP}$\rem{compliant with} \add{following from} the Hagen-Pouseuille law. Moreover, it shows the impact of the tortuosity $\tau \approx 3.6$. A value of $\tau$\rem{in the proximity of} \add{of about} three seems reasonable if one considers that in an isotropic medium such as Vycor, the porosity can, to first approximation, be accounted for by three sets of parallel capillaries in the three spatial directions; but only one third of these capillaries sustain the flow along the pressure drop. A value larger than three reflects the extended length of a meandering capillary beyond that of a straight one.

Following these preliminary considerations, one may now take the leap to the imbibition \add{kinetics}\rem{dynamics}. In general, there are three characteristic regimes during a capillary rise process \cite{Bonn2009, Courbin2009, Oyarzua2015}: (i) The initial, ballistic regime when the fluid particles enter the pore. In this region, the transport is governed by competition between inertia and molecule/surface interactions resulting in a linearly increasing capillary rise height \cite{Bosanquet1923, Quere1997, Kornev2001}. (ii) The regime where inertia and viscous forces compete with capillarity \cite{Bosanquet1923, Kornev2001}. (iii) The regime of viscous flow, where viscous forces prevail and act against the capillarity-driven liquid uptake by the porous medium \cite{Bell1906, Lucas18, Washburn21}. For the time resolution of the present experiments, only regime (iii) is of relevance. The typical time $\tau_{\rm int}$ that is required for the viscous flow to dominate in the pore can be estimated with \cite{Gennes2004} $\tau_{\rm int} =\rho r_0^2/\eta$. This typically yields times on the order of some 10 ps.

In the viscosity-dominated regime of the capillary rise, the liquid encounters the pressure drop $\Delta p$ along a time-dependent length $h(t)$, equivalent to the actual rise height, meaning $d\to h(t)$ -- see Fig.~\ref{fig1} for a schematic illustration. Moreover, with the initial porosity $\phi_{\rm i}$ of the sample the permeated volume can be determined at any time $t$ to be $V(t) =  \phi_{\rm i} \, A\, h(t)$. Accordingly, Eq.~\eqref{eq:Darcyslaw} becomes a simple differential equation
\begin{equation}
  V\; \dot{V} = A^2 \; \frac{\phi_0 \, \phi_{\rm i} \, r_{\rm h}^4 }{8 \, r_0^2 \, \tau \, \eta} \; \Delta p \; .
\label{eq:imbDE}
\end{equation}
Equation \eqref{eq:imbDE} is solved by a $\sqrt{t}$ law for $V(t)$ \cite{Lucas18, Washburn21} to yield the rise height $h(t)$
\begin{equation}
h(t) = \underbrace{\sqrt{\frac{\sigma\, \cos\theta}{2\,\phi_{\rm i} \, \eta}} \; \Gamma}_{C_{\rm h}} \; \sqrt{t} 
\label{eq:LWht}
\end{equation}
as well as the sample's mass increase $m(t)$ due to the liquid uptake $m(t)$\add{,} 
\begin{equation}
m(t) = \underbrace{ \rho \; A\;\sqrt{\frac{\phi_{\rm i} \,\sigma\, \cos\theta}{2\, \eta}} \; \Gamma}_{C_{\rm m}} \; \sqrt{t} \; .
\label{eq:LWmt}
\end{equation}
Assuming the liquid bulk parameters to be still valid under mesopore confinement, following Eq. \ref{eq:imbDE} and \ref{eq:LWht} the proportionality constant becomes:
\begin{equation}
\Gamma = \frac{r_{\rm h}^2}{r_0} \;\sqrt{\frac{\phi_0}{\tau\,r_{\rm L}}}
\label{eq:Gamma}
\end{equation}
Thus, $\Gamma$ can be inferred from both rise height ($h(t) \to C_{\rm h} \to \Gamma$) and mass increase ($m(t) \to C_{\rm m}\to \Gamma$) measurements. As $\Gamma$ is determined by only matrix-specific quantities (and in particular no liquid properties) it should be a constant for matrices with identical internal structure and chemical composition -- independent of the imbibed liquid. Therefore we will refer to $\Gamma$ as \textit{imbibition ability}. The larger $\Gamma$ the faster is the imbibition process. The unit of $\Gamma$ is inverse square-root of length. It is direct\add{ly} proportional to the square root of the pore dimensions expressed by a pore radius $r$ (as can be easily seen by applying $ r_{\rm h}= r_{\rm 0}= r_{\rm L}\equiv r $). Hence, the liquid will rise \add{faster in larger pores}\rem{the faster the larger the pores are}.

The concept of the imbibition ability enables a direct comparison of measurements with different liquids. Measuring the homologous series of n-alkanes\add{,} this approach should reveal any influence of the chain length $\ell$ as $\frac{{\rm d}\Gamma}{{\rm d}\ell} \ne 0 $. Furthermore, its absolute value contains\rem{many} information on the nanoscopic flow behaviour, especially on the hydrodynamic radius $r_{\rm h}$ and hence on the exact hydrodynamic boundary condition. Thus, our measurements provide an ideal method for studying the liquid dynamics in the proximity of interfaces as well as in extreme confinement. 

\section{Materials and methods}


\begin{figure} \center
\includegraphics[width=.95\columnwidth]{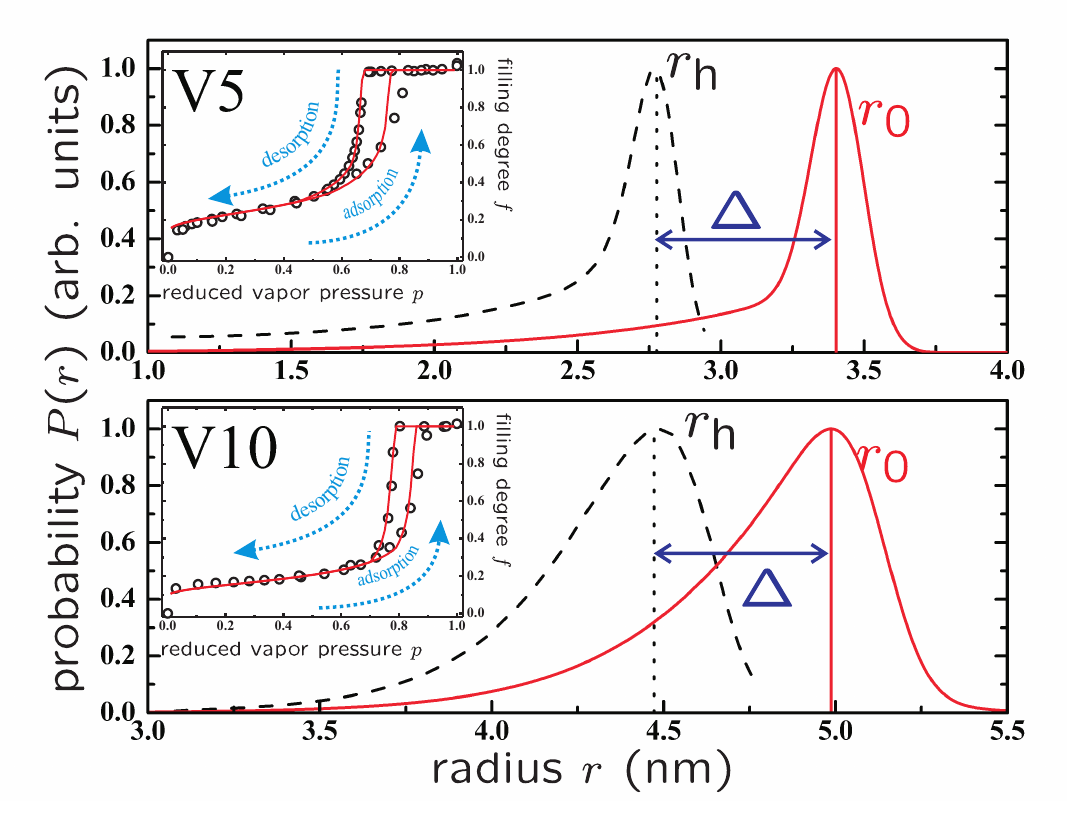}
\caption{(color online). Distributions of pore radii (solid lines) as extracted from nitrogen sorption isotherm measurements and  hydrodynamic pore radii (dashed lines) as extracted from neutron radiography measurements for both V5 (upper panel) and V10 (lower panel). The peak values of the pore radius distribution, $\overline{r}_0$ (3.40~nm and 4.99~nm), and the hydrodynamic pore radius distribution, $\overline{r}_{\rm h}$ (2.77~nm and 4.47~nm), are indicated by the dotted lines, $\Delta$ denotes the peak shift (0.63~nm and 0.52~nm). Insets: Nitrogen sorption isotherms of V5 and V10 recorded at 77~K. Shown is the filling $f$ versus the reduced vapor pressure $p$. The data points with $f>1$ indicate the formation of bulk liquid droplets outside the mesopores. Solid lines are calculations within a mean field model. They are based on the pore radius distributions shown as solid lines in the main plots.}
\label{fig2_3}
\end{figure}

\subsection{Porous glass substrates}
The spatial restrictions in the nanometer range were provided by the sponge-like topology of porous Vycor glass (Corning, code 7930). Vycor is virtually pure fused silica glass permeated by a three-dimensional network of interconnected pores \cite{Levitz91, Mitropoulos95, Gelb98, Huber1999}. The experiments were performed with two types of Vycor\rem{significantly} differing in the mean pore radius $\overline{r}_0$ only.  Their volume porosity is $\phi_0 \approx 0.3$.\rem{The aspect ratio $a$=pore diameter/ pore length of Vycor glasses is between 5 and 7 \cite{Levitz91, Mitropoulos95, Gelb98}.} For convenience the two types will be termed V5 ($\overline{r}_0=3.4$~nm) and V10 ($\overline{r}_0=5.0$~nm) in the following. The pore shape in Vycor glasses is to a first approximation 'cylindrical' with an aspect ratio, i.e. ratio of pore diameter to pore length, larger than 1 as has been inferred from electron micrographs by an analysis of chord length distributions \add{\cite{Levitz91, Mitropoulos95, Gelb98}}. 

The\rem{just mentioned} matrix properties were determined by nitrogen sorption isotherms performed at 77~K. The isotherms are shown as insets in Fig.~\ref{fig2_3}. We plot the filling $f$, that is, the number of nitrogen molecules adsorbed by the matrix normalized by the amount of nitrogen necessary for its complete filling, versus the reduced vapor pressure $p=p_{\rm e}/p_0$. The pressure $p_0$ refers to the bulk vapor pressure of nitrogen at $T=77$~K and $p_{\rm e}$ to the equilibration pressure after each adsorption or desorption step\rem{, respectively}. 

In a subsequent step, the isotherms were analyzed within a mean field model proposed by Saam and Cole \cite{Saam75} for cylindrical pore geometry. Using a trial-and-error technique, we calculated isotherms corresponding to preset pore size distributions $P(r)$ and compared the results with our measurements. In several \add{iterations}\rem{loops} the initial distributions were adjusted until calculations and experiments coincided adequately. The\rem{adapted} distributions obtained are shown as solid lines in Fig.~\ref{fig2_3}, the associated isotherms are indicated by the solid lines in the insets of Fig.~\ref{fig2_3}. Both Vycor samples seem to have\rem{a most probable pore radius $r_0$ (as indicated by the dotted lines) surrounded by} an asymmetric distribution of \add{pore radii with} several larger but far more smaller pores. 

Apart from the pore radii the capillary rise dynamics are\rem{additionally} \add{also} influenced by the network morphology \cite{Cai2011, Cai2014}. The simplest analytical approach in order to account for the spongelike structure of the sample (see Fig.~\ref{fig1}\add{(a)}\rem{(b)} for an illustration) is the introduction of the so-called tortuosity $\tau$. It can be inferred from experiments on the self-diffusion of liquids in porous Vycor compared to the fluid's bulk kinetics \rem{behaviour}. Small angle neutron scattering measurements of diffusion coefficients \cite{Lin92} resulted in $\tau =3.6 \pm 0.4$. This value is in good \add{agreement}\rem{accordance} with simulations of Vycor's pore morphology \cite{Crossley91}.

We cut regularly shaped blocks of height $d$ ($\sim 20$~mm) from the delivered rods. Prior to \add{use}\rem{using}, we subjected them to a cleaning procedure with hydrogen peroxide and nitric acid followed by rinsing in deionized Millipore water and drying at 200~\degr C in vacuum for two days. This treatment ensures \add{that}\rem{the removing of} any organic contamination \add{is removed from}\rem{on} the large internal surface of the samples. Until \add{use}\rem{usage}, the samples were stored in a desiccator. 

\subsection{Liquid hydrocarbons}
The Vycor matri\add{ces} provide spatial restrictions of only a few nanometers, so they are\rem{solely} \add{only} ten to 100 times larger than the typical molecular dimensions of simple liquids. Hence, it is not obvious that the standard assumptions of continuum fluid mechanics hold in a system where no more than 1000 molecules per cross-sectional area can be accommodated. To clarify this, we performed a systematic study of the flow dynamics in confinement as a function of the underlying complexity and structure of the liquid's building blocks. 
\begin{table*}[!t]
\centering
\caption{Densities $\rho$, viscosities $\eta$, surface tensions $ \sigma$, contact angles $\theta$ with silica and all-trans molecule lengths $\ell$ for the\rem{used} n-alkanes \add{studied} at the temperature $T$ of the measurements. The values are taken from: \mk{a} \cite{Small86}, \mk{b} \cite{LB_IV_8B}, \mk{c} \cite{LB_IV_18B}, \mk{d} \cite{LB_IV_16}. Contact angles were\rem{in-house determined to be} well below 10\degr. The info-column contains information on the\rem{respective} supplier\add{s} ({\sc AA}: Alfa Aesar, {\sc Al}: Aldrich, {\sc Fl}: Fluka, {\sc Me}: Merck) and the alkane's minimum purity (in percent). \mk{s} squalane (2,6,10,15,19,23-hexamethyltetracosane). } 
\begin{tabular}{C{\linewidth/9}*{7}{|C{\linewidth/9}}}
 alkane & info & $T$ (\degr C)   & $\rho$ (g/ml)   &  $\eta$ (mPa$\,$s)  &  $\sigma$ (mN/m)  & $\cos\theta$ &$\ell$ (nm)  \\ \hline \hline
             n-\ce{C10H22}     & {\sc Me}99                  & 25 & 0.7271\mk{a} &  0.8835\mk{a} &  23.49\mk{a} & 1 & 1.34 \\ 
             n-\ce{C12H26}     & {\sc Me}99                  & 25 & 0.7466\mk{a} &  1.381\mk{a}  &  25.10\mk{a} & 1 & 1.59 \\ 
             n-\ce{C14H30}     & {\sc AA}99                  & 25 & 0.7595\mk{a} &  2.086\mk{a}  &  26.21\mk{a} & 1 & 1.84 \\ 
             n-\ce{C16H34}     & {\sc Me}99                  & 25 & 0.7681\mk{a} &  3.087\mk{a}  &  26.97\mk{a} & 1 & 2.09 \\ 
             n-\ce{C18H38}     & {\sc Fl}99                  & 36 & 0.7711\mk{a} &  3.407\mk{a}  &  26.88\mk{a} & 1 & 2.34 \\ 
             n-\ce{C20H42}     & {\sc Al}99                  & 41 & 0.7748\mk{b} &  4.06\mk{c}   &  27.12\mk{d} & 1 & 2.59 \\ 
             n-\ce{C24H50}     & {\sc Fl}99                  & 54 & 0.7768\mk{b} &  4.948\mk{c} &  29.78\mk{d} & 1 &  3.09 \\ 
             n-\ce{C30H62}     & {\sc Fl}98                  & 74  & 0.7765\mk{b} &  5.47\mk{c} &  26.56\mk{d} & 1 &  3.84 \\  
             n-\ce{C40H82}     & {\sc AA}97                  & 91  & 0.7748\mk{b} &  7.20\mk{c} &  26.04\mk{d} & 1 &  5.09 \\  
             n-\ce{C60H122}    & {\sc Fl}98                  & 107 & 0.7720\mk{b} &  13.14\mk{c} &  24.65\mk{d} & 1 & 7.59  \\ 
               \ce{C30H62}\mk{s}   & {\sc Me}99                  & 30 & 0.8016\mk{b} &  22.13\mk{c} &  28.14\mk{d} & 1 &       \\ 
\end{tabular}
\label{liquidprops}
\end{table*}

\rem{At this} The homologous series of linear hydrocarbons (n-alkanes, n-C$_{n}$H$_{2n+2}$) is\rem{rather} well \add{suited}\rem{applicable} for this purpose. By\rem{simply} extending the carbon backbone of the alkane one can\rem{easily} change the molecule's dimensions whilst all other liquid properties (density $\rho$, surface tension $\sigma$, viscosity $\eta$, contact angle with silica $\theta$)\rem{nearly} remain almost constant -- see Tab.~\ref{liquidprops} for details. In particular, all alkanes completely wet silica\rem{what} \add{which} is the basic condition for spontaneous imbibition measurements in Vycor glass. For the main part of this study we used linear hydrocarbons ranging from n-decane (n-\ce{C10H22}) to n-hexacontane (n-\ce{C60H122})\rem{being tantamount} \add{corresponding} to all-trans molecule lengths between 1~nm and approximately 8~nm. In addition, the branched alkane squalane (2,6,10,15,19,23-hexamethyltetracosane) was also \add{studied}\rem{used} in order to\rem{scrutinize impacts} \add{investigate the effects} of the molecule's internal structure as opposed to its\rem{sheer} length. 

\subsection{Experimental techniques}
In this study, we combine three different measurement methods. The capillary rise dynamics of n-alkanes in porous Vycor glass have been recorded by using gravimetrical, optical as well as neutron radiography measurements. 

\subsubsection{Gravimetrical measurements}
\begin{figure} \center
\includegraphics[width=.8\columnwidth]{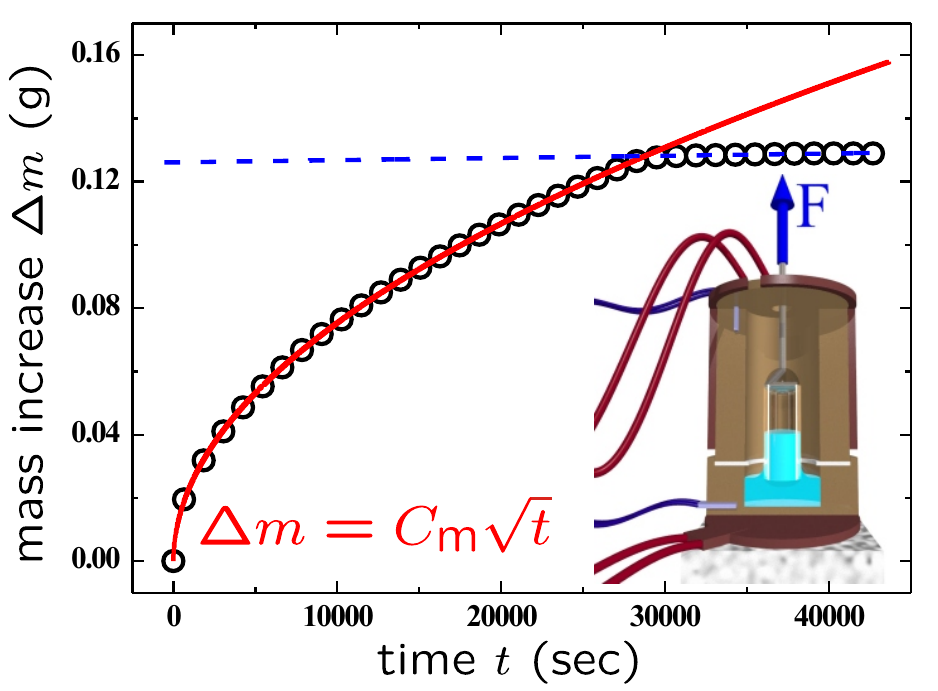}
\caption{(color online). Typical measurement of the increase in mass (circles) of a porous Vycor block (V10) due to the imbibition of a liquid (n-tetradecane) at room temperature. According to Eq.~\eqref{eq:LWmt} the prefactor of the\rem{shown} $\sqrt{t}$-Lucas-Washburn kinetics fit (solid line)\rem{comprises} \add{provides} information on the microscopic flow behavior expressed in terms of the imbibition ability $\Gamma$. The\rem{gradual} mass increase comes to a halt and a constant plateau (dashed line) is reached, when the sample is completely filled. For clarity \add{only every 1200th data point is shown}\rem{ the data density was reduced by a factor 1200}. Inset:\rem{Raytracing} Illustration of the thermostated imbibition cell\rem{employed} \add{used} for the isothermal capillary rise experiments.}
\label{imb_bsp}
\end{figure}

According to Eq.~\eqref{eq:LWmt} the imbibition dynamics and hence the imbibition ability $\Gamma$ can be\rem{easily} inferred from the increase in the sample's mass due to liquid uptake during the imbibition experiment. Such {\it gravimetrical measurements} can be easily performed by means of a standard laboratory balance. The sample is attached to the balance with a special mounting,\rem{so} \add{hence} allowing time dependent recording of the gravitational force acting on the porous block. A surrounding copper cell enables thermostating of both the sample and the liquid reservoir \cite{Gruener2009a} -- see inset in Fig.~\ref{imb_bsp} for\rem{a raytracing} \add{an} illustration. The measurement is started by moving the reservoir upwards until the liquid surface touches the bottom of the porous host. The rise process starts immediately after the contact. 

\subsubsection{Optical imaging}
In case of room temperature measurements that do not require the thermostating copper jacket, the capillary rise process has also been followed {\it optically}\rem{recorded} by means of a CCD monochrome camera placed in front of the setup. At this point one might wonder whether we observed any optical signal. This question is justified as there is indeed no visual difference between the completely filled and the completely empty part of the sample. This is because the \add{length scale of the} spatial variations in the refractive index\rem{being tantamount} is far too small \add{due} to the pore-pore distances of only\rem{some} \add{a few} nanometers. Therefore, visible light is neither scattered in the completely empty nor in the completely filled parts of the block. Hence, the measurement of the actual rise height and a subsequent calculation of the imbibition ability $\Gamma$ via Eq.~\eqref{eq:LWht} is not possible.

\begin{figure*} \center
\includegraphics[width=.95\linewidth]{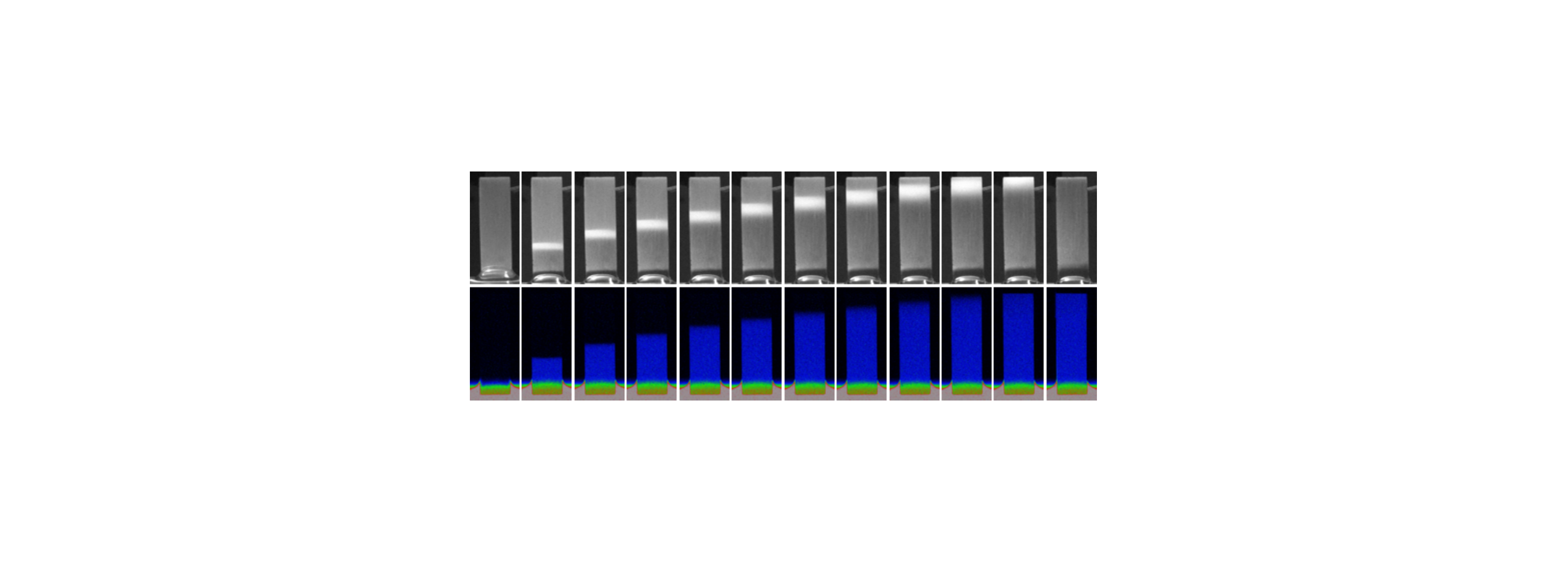}
\caption{(color online). n-tetradecane invading a Vycor block (V10, $V=4.55\times 4.55\times 15.00$~mm$^3$) visualized by (\rem{1st} \add{top} row) a series of \add{images}\rem{pictures} taken by means of a CCD monochrome camera and (\rem{2nd} \add{bottom} row) a series of processed radiography\rem{pictures} \add{images, which are}\rem{. The highlighting of variances is enhanced by rendering} \add{rendered}\rem{the radiography images} in pseudocolors. The time interval between\rem{each picture} \add{subsequent images} is approximately 50 min\rem{utes}. }
\label{c14_v10}
\end{figure*}

Nevertheless, during the imbibition, an opaque front moves from the bottom of the sample to its top, where it finally vanish\add{e}s. During this process the front's width gradually increases -- see\rem{first} \add{top} row in Fig.~\ref{c14_v10}. Following the above argument, the \add{region}\rem{area} of the block showing scattering can be neither completely filled nor completely empty. It contains regions of filled \add{pores}\rem{paths} as well as still empty ones, resulting in variations of the refractive index on length scales including the wavelength of visible light.\rem{The so induced light scattering is} \add{This leads to the scattering of light,} a phenomenon well known from the emptying of completely filled Vycor samples via percolating paths \cite{Page93, Ogawa2013, Ogawa2015} or collective rearrangements of pore condensate induced by phase transitions \cite{Soprunyuk2003, Soprunyuk2004, Huber2013, Varanakkottu2014} in porous glass. Its existence in this context shows that there is no sharp interface that separates the filled from the still empty sample, but an extended front separating the two regions. Unfortunately, the location and the width of the front can not easily be resolved by means of optical measurements alone as strong multiple scattering makes its analysis and hence the quantitative determination of the exact liquid distribution within the light\add{-}scattering front \add{impossible}.\rem{For this purpose} \add{Thus} we\rem{harnessed} \add{applied} a different method, neutron radiography, that\rem{is sensitive to} \add{yields} the local liquid concentration\rem{ and not only to special structural arrangements of the fluid}.

\subsubsection{Neutron radiography}

{\it Neutron radiography} is an ideal method to observe the imbibition dynamics in porous Vycor glass. This is because of the weak interaction of thermal neutrons with silica (total scattering cross sections $\sigma_{\rm tot}\approx 2.3$~barn) and oxygen ($\sigma_{\rm tot}\approx 4.2$~barn) as opposed to hydrogen ($\sigma_{\rm tot}\approx 82$~barn). Therefore it is easy to \add{perform}\rem{do} time-resolved quantitative \add{measurements}\rem{analyses} of the distribution of hydrocarbons within the almost transparent silica matrix. Furthermore, using aluminium ($\sigma_{tot} \approx$ 1.7 barn), sample cells and holders which only weakly interact with neutrons can be constructed.The experiments were performed at the ANTARES beamline of the Heinz Maier-Leibnitz (FRM~II) research reactor of Technical University Munich in Garching, Germany \add{\cite{Schillinger2006}}.
\begin{figure} \center
\includegraphics[width=.8\columnwidth]{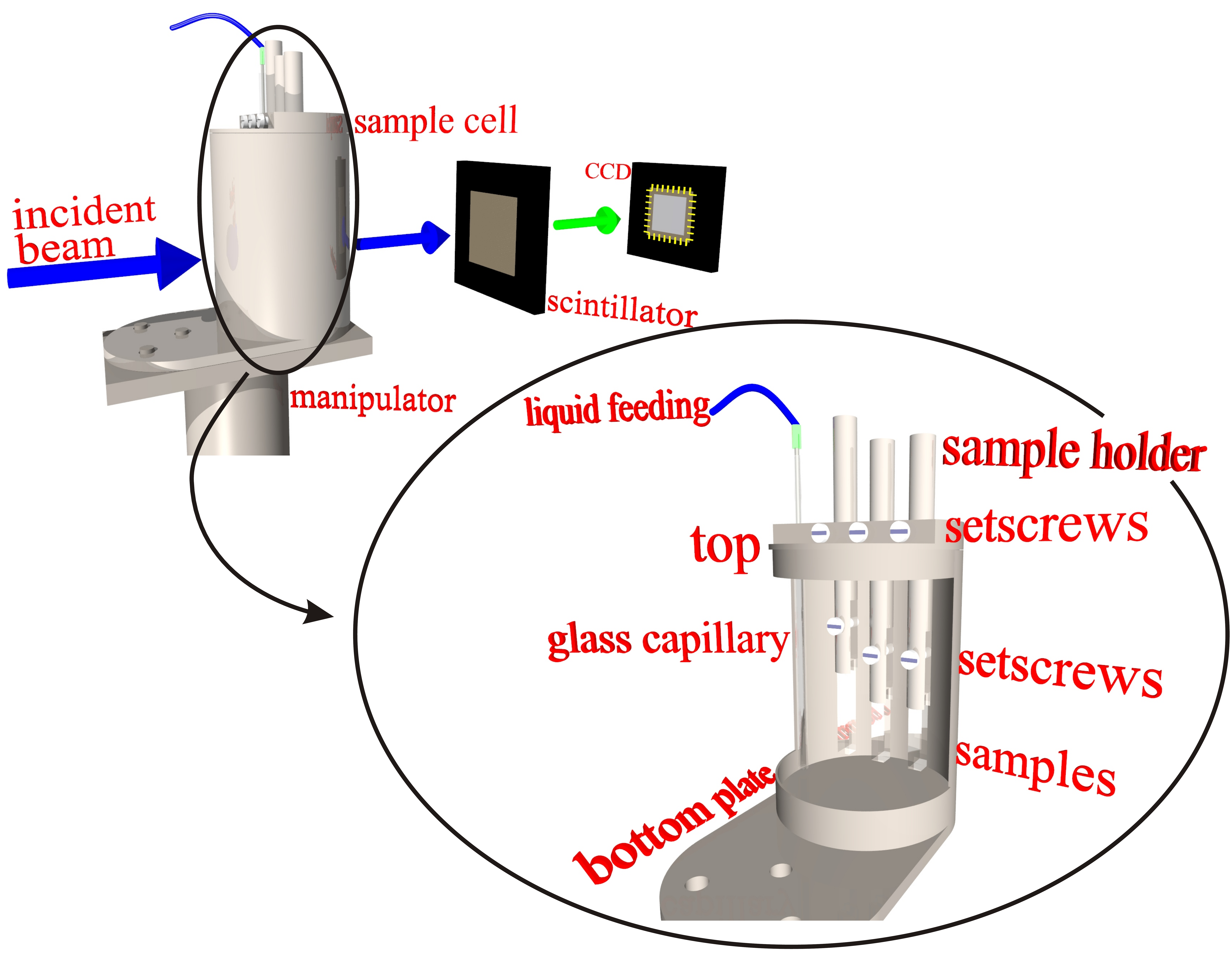}
\caption{(color online).\rem{Raytracing} Illustration of the\rem{structural principle of the} neutron radiography setup\rem{ANTARES}. Blue arrows indicate neutron radiation, the green arrow indicates visible light. The magnified\rem{image} detail\rem{gives} \add{provides} a view \add{of the} inside \add{of} the sample cell. The whole cell and the sample holder are made of aluminium\add{, which only very weakly interacts}\rem{because of its weak interaction} with neutrons. The samples are mounted on the clamp-like sample holders. Up to three \add{samples}\rem{of them} can be simultaneously attached to the cell. The liquid is supplied via a \add{tube}\rem{lead} and a glass capillary and driven by means of an external pump.}
\label{NRS}
\end{figure} 

The general design of \add{a neutron radiography experiment}\rem{ANTARES} and a detailed view\rem{into} \add{of} the sample cell is illustrated in Fig.~\ref{NRS}. The incident beam of\rem{moderated} \add{cold} neutrons has a maximum divergence of 0.3$^\circ$. At the location of the specimen the flux is approximately $10^8~{\rm s}^{-1}{\rm cm}^{-2}$. After penetrating the sample cell\rem{with} \add{and} the samples, the attenuated neutron beam hits a ZnS+LiF detector for the space \add{resolved}\rem{dispersive} detection of the neutron intensity. The emitted visible light is\rem{finally} recorded by a camera with a CCD chip with $2048\times 2048$ pixel, which is - at the selected lens magnification - equivalent to a pixel size in the sample plane of $\sim 16$~$\upmu$m, giving a field of view of 32.7 $\times$ 32.7~cm$^2$. The measurement is started by pumping the liquid into the\rem{previously} \add{initially} empty sample cell until the liquid surface touches the bottom face\rem{t}s of the samples. For each\rem{taken picture} \add{image} the CCD chip is exposed for a constant time interval of 30 s\rem{econds}. This \add{measurement} time\rem{scale is} \add{represents} a compromise between sufficient statistics and \add{the best} time resolution \add{making sure that no structural detail moved further than the spatial resolution thus avoiding smearing of the concentration profile of the liquid}. 

In the course of the\rem{subsequent evaluation} \add{analysis} the obtained transmission\rem{pictures} \add{images} were all\rem{normalized with respect to} \add{corrected using} a reference\rem{picture} \add{image} of the completely empty sample cell taken prior to \add{the} measurement \add{of the sample}\rem{start} \add{and normalized to the incident beam intensity and detector efficiency using an image of the direct beam}. A dark frame containing the constant camera offset and the dark noise signal has been subtracted from all images. The\rem{residual} \add{remaining} signal is only caused by the liquid sucked in the sample. Therefore, the systematic evaluation of such\rem{photographs} \add{images} as a function of time provides detailed information about the imbibition dynamics. A series of such \rem{pictures} \add{images} is shown in the\rem{second} \add{bottom} row of Fig.~\ref{c14_v10} for n-tetradecane\rem{being sucked} \add{imbibing} into a V10 sample \footnote{Note that the same images were published erroneously in \cite{Gruener2012}, which reports on the imbibition of water, although the images show the imbibition of n-tetradecane.}. In the processed \rem{pictures} \add{images}, filled areas are coloured whereas still empty parts are not visible. At first sight they\rem{may} resemble the optical set of\rem{photographs} \add{images} in the\rem{first} \add{top} row of Fig.~\ref{c14_v10}. But there are basic differences in the\rem{comprised} information contained: As mentioned previously the optical measurements only reveal qualitative information about the location of partly filled areas. On the other hand, neutrons are scattered by hydrogen, which has nearly equal mass, into 4 $\pi$ space. Thus, almost all scattered neutrons are removed from the path to the detector and hence the direct beam is attenuated, which appears effectively like absorption. Only a small part of the neutrons is scattered under small angles and reaches the detector as a diffuse background, which can be neglected for small amounts of hydrogen. Therefore, the attenuation in the radiography \rem{pictures} \add{images}\rem{correspond} is directly \add{related} to the local\rem{fluid} \add{liquid} concentration and, therefore, allows for a\rem{thorough} \add{detailed quantitative} analysis of the imbibition dynamics and the advancing liquid front based on the\rem{actual} liquid distribution within the sample at any given time.

\section{Results}
\subsection{Imbibition ability as a function of molecule length and humidity}
\begin{figure} \center
\includegraphics[width=.65\columnwidth]{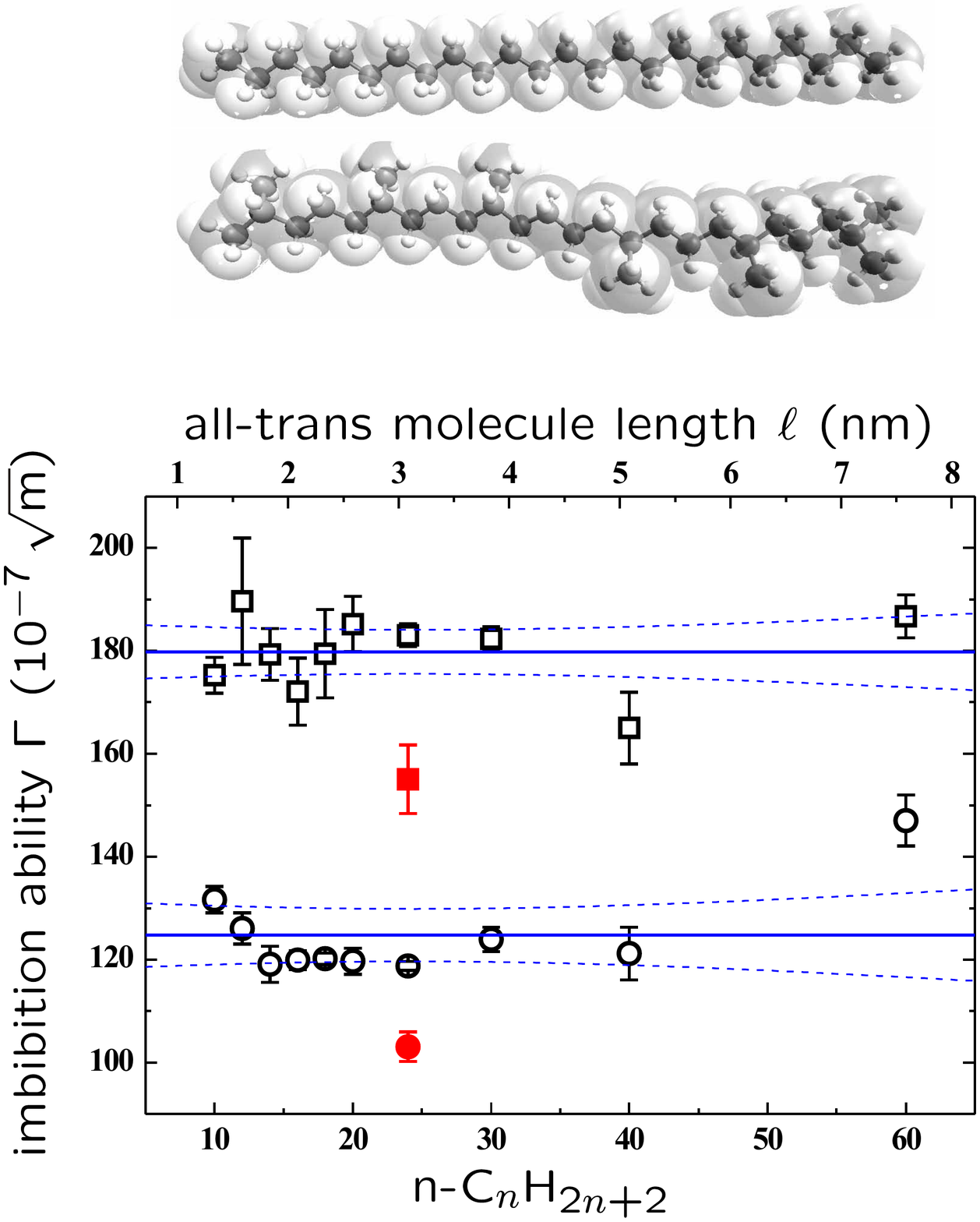}
\caption{(color online). (\rem{upper panel} \add{top})\rem{Illustrations} \add{Schematic representations} of the all-trans configurations of the linear hydrocarbon n-tetracosane ($n$-C$_{24}$H$_{50}$) and the branched hydrocarbon squalane ($n$-C$_{30}$H$_{62}$). The backbones of both chains contain 24 C atoms, so they\rem{coincide in the} \add{have about the same} all-trans molecule length $l$ $\approx$ 3 nm. (\rem{lower panel} \add{bottom}) Imbibition abilities $\Gamma$ obtained from gravimetrical measurements for a series of linear hydrocarbons (open symbols) ranging from n-decane to n-hexacontane invading into V10 (squares) and V5 (circles), respectively. Their mean values are visualised by the solid lines, lying within the confidence interval (dashed lines, confidence level: 90~\%). The corresponding all-trans molecule length $\ell$ of the alkane is indicated by the top axis. The\rem{measured} \add{data of the} branched hydrocarbon squalane\rem{is} \add{are} represented by the full symbols.}
\label{conductance}
\end{figure}

For every alkane listed in Tab.~\ref{liquidprops} we performed gravimetrical measurements in V5 and V10. Each liquid/substrate combination was measured at least three times. From the $\sqrt{t}$-fits \add{to}\rem{of} the obtained mass increase \add{data}\rem{curves} (as shown in Fig.~\ref{imb_bsp}) we extracted the proportionality constant $C_{\rm m}$. In combination with Eq.~\eqref{eq:LWmt} this\rem{finally} yields the imbibition ability $\Gamma$. The results\rem{of this study} are shown in Fig.~\ref{conductance} as a function of all-trans molecule length $\ell$ for both V5 and V10. 

\rem{With respect to the results shown in} Fig.~\ref{conductance}\rem{it is clearly recognizable} \add{shows} that the measured imbibition abilities can be well described by the arithmetic averages over all measured linear alkanes\add{;} $\Gamma=(124.8\pm	2.8)\times 10^{-7}~\sqrt{\rm m}$ for V5 and $\Gamma=(179.8 \pm	2.3)\times 10^{-7}~\sqrt{\rm m}$ for V10, respectively. Over a wide range of molecule lengths $\ell$ there is no\rem{influence} \add{effect} on the imbibition kinetics. This statement is corroborated by the\rem{inserted} confidence intervals that broaden only slightly\rem{and merely} for the highest $\ell$s and, hence, suggest that\rem{the `true'} $\Gamma$ is\rem{obtained in an} $\ell$-invariant\rem{interval}. Apart from the squalane experiments a significant deviation can only be observed for n-hexacontane (n-\ce{C60H122}) in V5 \add{which will be discussed below}. 

In addition, using a humidifier, measurements of n-hexadecane (n-\ce{C16H34}) invading into V5 were performed as a function of relative humidity\rem{ in the laboratory}. The resultant imbibition abilities are listed in Tab.~\ref{gammaC16} together with the extracted values of the initial porosities $\phi_{\rm i}$. No distinct influence of the humidity on the imbibition dynamics, except for the porosity changes, is detectable.

\begin{table}
\centering
\caption{Imbibition abilities $\Gamma$ of n-hexadecane invading into V5 as a function of the relative humidity (RH)\rem{ in the laboratory}. In addition the initial porosities $\phi_{\rm i}$ as extracted from the mass increase measurements are listed.} 
\begin{tabular}{C{\columnwidth/4}|C{\columnwidth/4}|C{\columnwidth/3}}
RH (\%) & $\phi_{\rm i}$ & $\Gamma$ ($10^{-7}~\sqrt{\rm m}$)   \\ \hline \hline
                   24   &  0.275    & $116.5\pm 3.8$          \\  
                   34   &  0.262    & $121.9\pm 5.3$        \\                     
                   50   &  0.245    & $121.9\pm 4.1$            \\                    
\end{tabular}
\label{gammaC16}
\end{table} 

\begin{figure} \center
\includegraphics[width=.8\columnwidth]{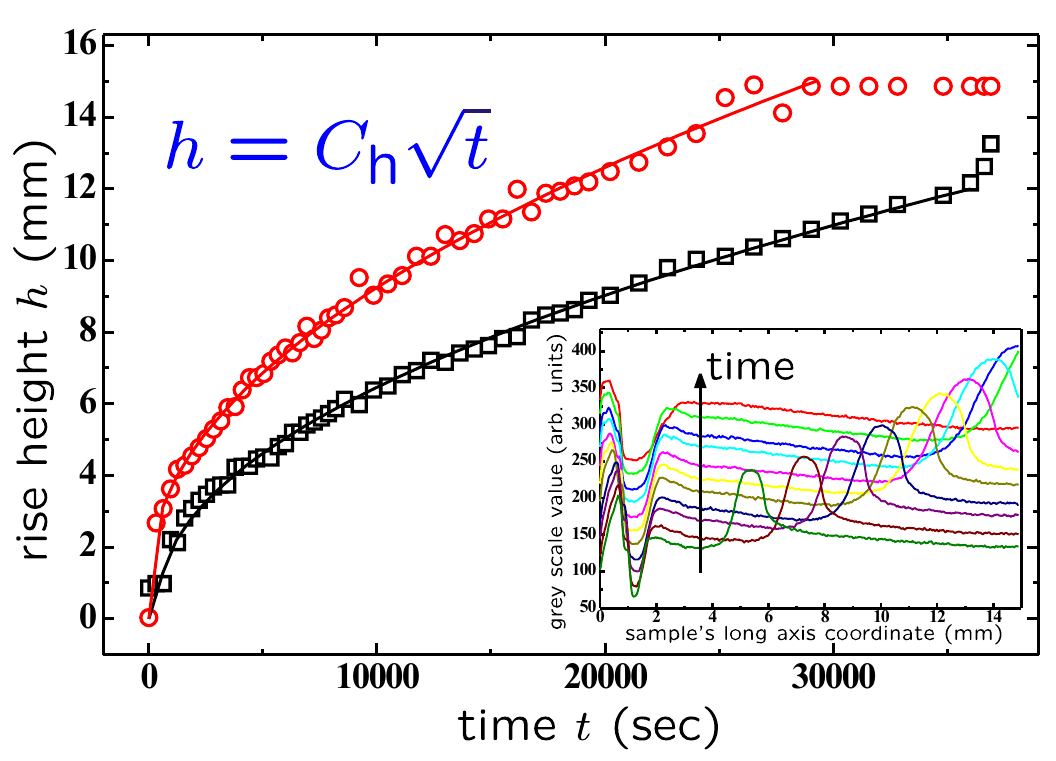}
\caption{(color online). Time dependent position of the upper (circles) and lower (squares) bounds of the light scattering front during imbibition of n-tetradecane in V5. Solid lines represent $\sqrt{t}$-fits to the data sets. The\rem{boundary's} positions are extracted from\rem{grey scale value} \add{transmission} profiles as\rem{exemplarily} \add{those} shown in the inset. For\rem{convenience} \add{clarity} each profile has been shifted upwards by a constant value.}
\label{c14_v10_optics}
\end{figure} 

\subsection{Imbibition front broadening as revealed by optical and neutron imaging}
{\rem{Additionally,} Imbibition experiments of n-tetradecane (n-\ce{C14H30}) and n-hexadecane (n-\ce{C16H34}, only in V5, but as a function of the relative humidity\rem{in the laboratory}) imbibing into porous Vycor were optically recorded in order to derive information on the light scattering front's position and width. The obtained series of pictures (e.g.\rem{first} \add{top} row in Fig.~\ref{c14_v10}) were \rem{systematically} evaluated\rem{by the help of an image processing program and a subsequent script-based analysis}. Based on time dependent\rem{grey scale value} \add{transmission} profiles we extracted the front's upper and lower bound, thereby gaining information on its position and width.

The behaviour of the front in particular the upper and lower bounds of the light scattering front are illustrated in Fig.~\ref{c14_v10_optics}. Comparing the results with the\rem{underlying} $\sqrt{t}$-fits it is clear that both bounds follow the same time dependency as the\rem{theoretical} rise height of a sharp imbibition front given by Eq.~\eqref{eq:LWht}. This suggests that there is not only {\it one} but rather a whole series of imbibition fronts causing the sample to fill inhomogeneoulsy\rem{what finally entails} \add{which leads to} the observed light scattering.\rem{In conjunction with the results of the neutron radiography measurements we will carry this discussion forward below. For convenience} From the front's upper and lower position, we have calculated $\Gamma$-values using Eq.~\eqref{eq:LWht}. This discussion will be continued later together with the neutron radiography results.

\begin{table}
\centering
\caption{Imbibition abilities $\Gamma$ (in $10^{-7}~\sqrt{\rm m}$)\rem{of} \add{based on} the\rem{light scattering} fronts' bounds as extracted from optical \add{measurements of the} rise height\rem{measurements} of n-tetradecane (C14) and n-hexadecane (C16) invading into V5 and V10\rem{, respectively}. The C16-measurements have been performed as a function of the relative humidity (RH\rem{, in \%})\rem{ in the laboratory}, but\rem{solely} \add{only} for V5. } 
\begin{tabular}{C{\columnwidth/8}|C{\columnwidth/8}|C{\columnwidth/8}||C{\columnwidth/4}|C{\columnwidth/4}}
\rem{V\ldots} \add{matrix} & fluid & RH \add{(\%)} & $\Gamma$ of upper bound  & $\Gamma$ of lower bound    \\ \hline \hline
 \multirow{4}{*}{V5}  &                C14     &              & 117.5            & 94.8  \\ \cline{2-5}
                     & \multirow{3}{*}{C16}   &   24         & 122.7            & 99.2  \\  
                     &                        &   34         & 122.4            & 90.5  \\                     
                     &                        &   50         & 119.9            & 84.3  \\  \hline                     
 V10                  &                C14     &              & 178.8            & 139.1  \\
\end{tabular}
\label{gammaheight}
\end{table}

The optical and gravimetrical results are now compared with the detailed\rem{concept of} \add{information on} the imbibition process obtained from the neutron radiography measurements.\rem{Thereby} We benefit from the sensitivity of\rem{the latter} \add{this} method to the local liquid concentration which results in\rem{rigorous and} time dependent information about the fluid distribution within the sample as visualized in the\rem{processed pictures} \add{images} in the\rem{second} \add{bottom} row of Fig.~\ref{c14_v10}.\rem{Grey scale value} \add{Transmission} profiles taken along the sample's long axis represent the basis of the subsequent analysis. They were normalized in such a way that they directly\rem{coincide with} \add{provide} the\rem{local} filling degree $0\le f \le 1$ of the sample\rem{at a certain} \add{as a function of the} height $h$ (see inset in Fig.~\ref{c14_v10_radiography}). 
\begin{figure} \center
\includegraphics[width=.8\columnwidth]{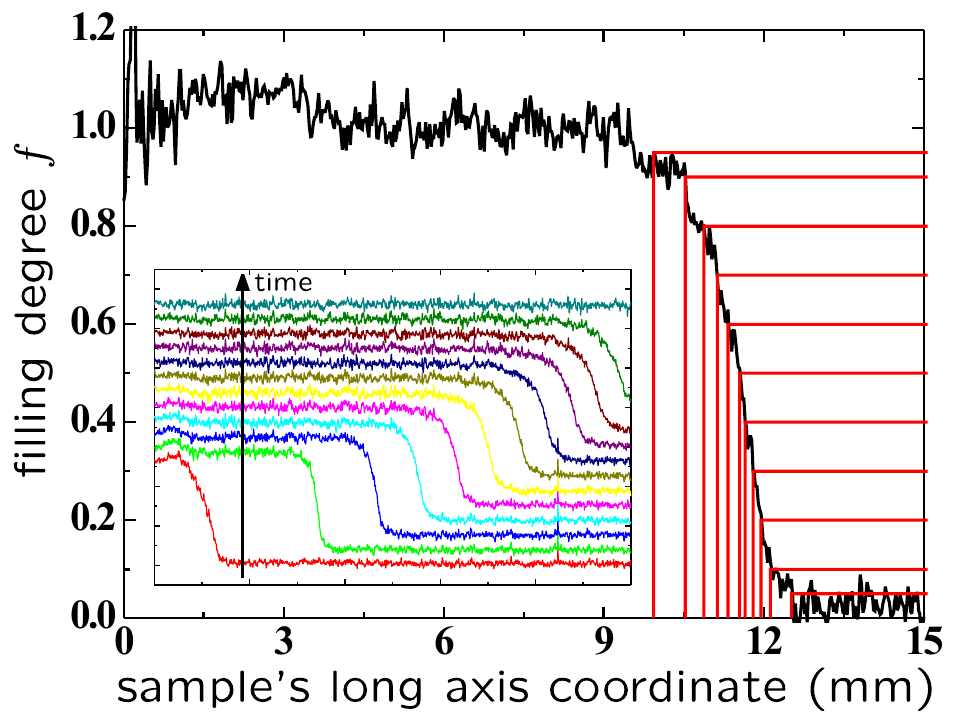}
\caption{(color online). Illustration of the method used to evaluate the normalized\rem{grey scale value} \add{transmission} profiles\rem{obtained from neutron radiography measurements} of n-tetradecane invading into \add{porous Vycor} V10 \add{obtained from neutron radiography measurements}. The\rem{shown grid} \add{horizontal lines} correspond to the\rem{following} filling degrees $f=5$~\%, 10~\%, 20~\%, \ldots, 80~\%, 90~\%, 95~\%. Note that the filling fraction fluctuations around 1, and in particular also values beyond 1, are characteristic of the error bars in the calculations of $f$ from the neutron imaging data. Inset: \add{Time} series of normalized\rem{grey scale value} \add{transmission} profiles of n-tetradecane invading into\rem{porous Vycor} \add{V10}. For\rem{convenience} \add{clarity} each profile has been shifted upwards by a constant value.}
\label{c14_v10_radiography}
\end{figure} 

Looking at the profiles shown in Fig.~\ref{c14_v10_radiography}(inset) it is clear that there is no sharp interface between the already completely filled ($f=1$) and the still empty part ($f=0$) of the sample. Comparable to the optical results\add{,} the advancing front exhibits \rem{a non-negligible} roughening. Hence, a thorough investigation of the profiles must\rem{cover} \add{include} the front's\rem{dynamics} \add{average position} as well as its shape\rem{'s behaviour}. For this purpose each profile was analysed by means of the grid method illustrated in Fig.~\ref{c14_v10_radiography}. For a series of filling degrees between zero and unity, we extracted the\rem{coordinate} \add{height} $h_f$ at which the\rem{respective level} \add{profile has a filling degree} $f$\rem{was traversed by the profile for the first time}. Repeating this procedure for all profiles,\rem{(} and therefore at different times\rem{ in the course of the experiment)}, results in a series of rise height curves $h_f(t)$.
\begin{figure} \center
\includegraphics[width=.8\columnwidth]{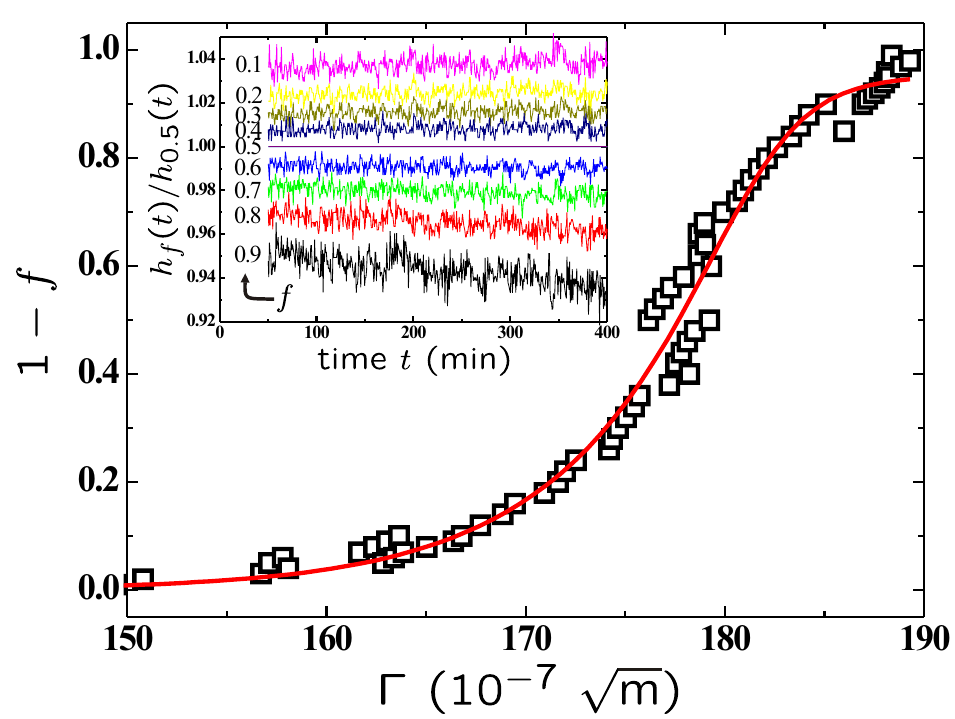}
\caption{(color online).\rem{Inverse filling degree} \add{Fraction of empty pore space,} $1-f$\add{,} as a function of the imbibition ability $\Gamma$\rem{(squares)} for n-tetradecane invading into V10. The solid line is a polynomial fit to the\rem{underlying} data \add{(squares)}\rem{set that is used for an analytical investigation of the $f(\Gamma)$-relation}. Inset: Selected rise height curves $h_f(t)$ \add{normalized by $h_{f=0.5}(t)$} for\rem{given} \add{different} filling degrees $f$\rem{normalized by $h_{f=0.5}(t)$} \add{(as indicated)}.\rem{The respective filling degrees are listed within the plot.}}
\label{c14_v10_radiography_2}
\end{figure} 

It is not\rem{too} surprising that all these rise height curves \add{$h_f(t)$} again obey a $\sqrt{t}$-law.\rem{This fact is verified by the $h_f(t)$-curves in the inset of Fig.~\ref{c14_v10_radiography_2}. Getting rid of} The $\sqrt{t}$-dependency \add{is taken into account} by\rem{normalizing them} \add{normalization} with the\rem{respective} \add{corresponding} measurement for $f=0.5$ \add{(inset of Fig.~\ref{c14_v10_radiography_2}).} It is evident that they all follow the same dynamic behaviour except for the\rem{$\sqrt{t}$-coefficient that is} time-independent \add{prefactor} $C_{\rm h}$, i.e. $C_{\rm h}=C_{\rm h}(f)$. Hence, the interface profiles all overlapp, provided one rescales the $h$-axis of the $h_f(t)$-curves by a transformation $h'_f=1/(C_{\rm h}(f) \sqrt{t})$. 

At this point we reconsider Eq.~\eqref{eq:LWht} and calculate imbibition abilities $\Gamma$ from the $C_{\rm h}$-values obtained from $\sqrt{t}$-fits to the rise height curves $h_f(t)$. Consequently, for each filling degree $f$ one\rem{ends up with} \add{obtains} a corresponding imbibition ability $\Gamma(f)$. The systematic implementation of this evaluation method provides detailed information on the imbibition dynamics,\rem{parametrised by} \add{namely} a close link between $\Gamma$ and $f$. This relationship is illustrated in Fig.~\ref{c14_v10_radiography_2}.

\begin{figure} \center
\includegraphics[width=.8\columnwidth]{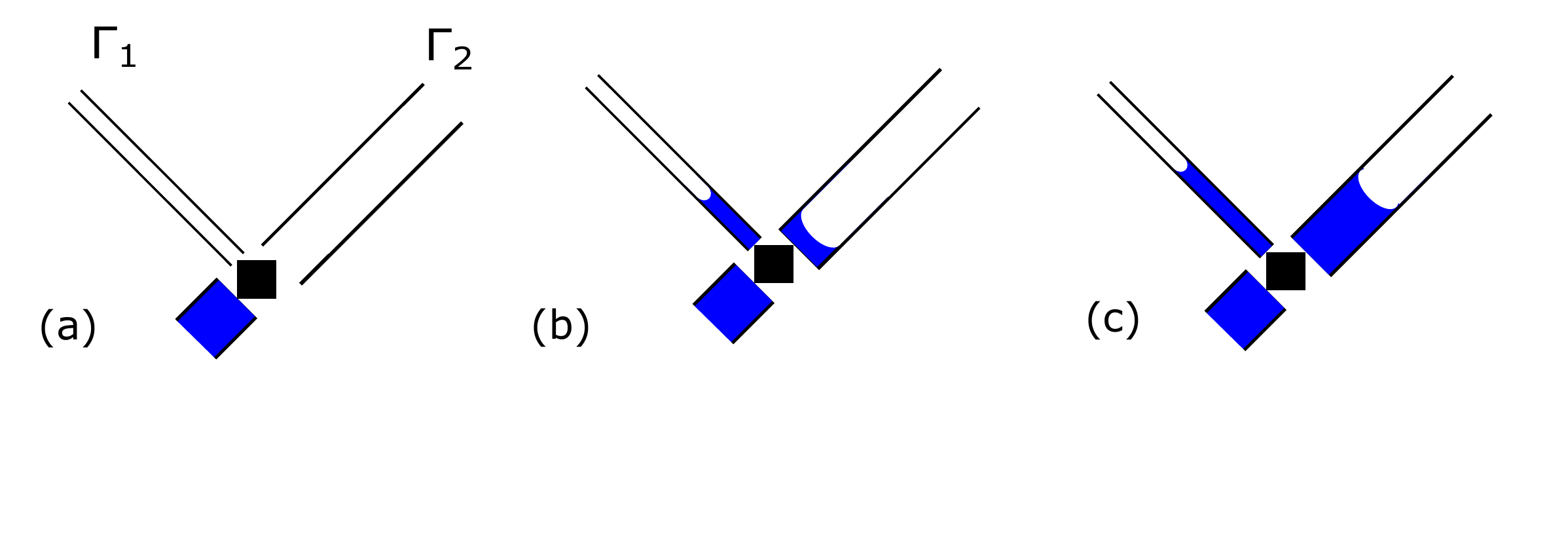}
\caption{(color online). Characteristic spontaneous imbibition events in a network of interconnected, elongated pores. (a) The advancing liquid reaches a pore junction with a large and small elongated pore and thus two distinct imbibition abilities $\Gamma_1<\Gamma_2$. At the junction two liquid menisci result which can move independently, coupled however by the hydrostatic pressure in the junction. In (b) a meniscus arrest in the large pore is illustrated, whereas the meniscus in the small pore continues to advance, as discussed in the text. In (c) both menisci move and the meniscus in the large pore advances faster than the one in the small pore, because $\Gamma_1<\Gamma_2$. The black squares represent the pore junction. Its filling kinetics and volume is considered as negligible.}
\label{fig:CRBroadening}
\end{figure} 

In order to emphasize the relationship's key information we plotted the\rem{inverse filling degree} \add{fraction of empty pore space,} $1-f$\add{,} as a function of $\Gamma$. The fraction $1-f$ can\rem{also} be\rem{seen} \add{regarded} as the\rem{integrated} \add{cumulative} probability to find a pore with an\rem{ny $\Gamma$ of maximum the corresponding} imbibition ability \add{smaller than} $\Gamma(1{-}f)$. This suggests that the filling process in Vycor results from a sequence of capillary filling events which is determined by the imbibition abilities of the porous medium. See Fig. \ref{fig:CRBroadening} for an illustration of filling events at a junction of an interconnected network of cylindrical capillaries with two radii, $r_1$ and $r_2$ (with $r_1<r_2$) and thus two distinct imbibition abilities ($\Gamma_1>\Gamma_2$). This conjecture will be addressed in more detail in the discussion part.

\begin{figure} \center
\includegraphics[width=.8\columnwidth]{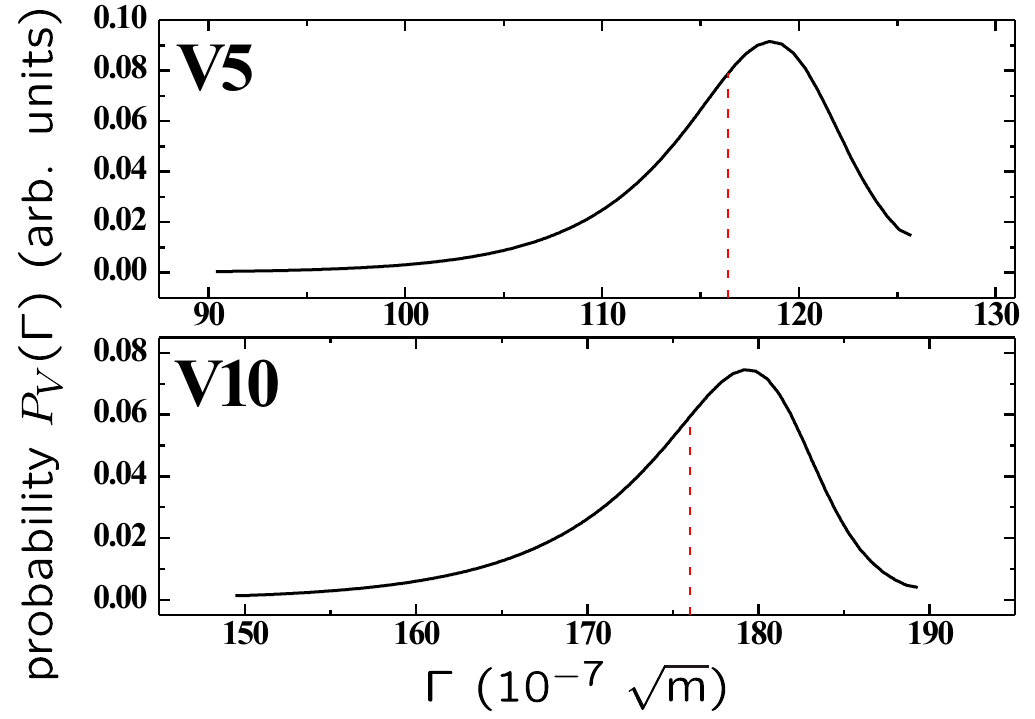}
\caption{(color online). (color online). Volume weighted probability distribution\rem{functions} $P_V(\Gamma)$ of the imbibition abilities $\Gamma$ for n-tetradecane invading into V5 (upper panel) and V10 (lower panel), respectively. The dashed lines indicate their volume weighted mean values $\overline{\Gamma}_V$.}
\label{c14_v10_radiography_3}
\end{figure}  

Since $(1-f)$ in Fig.~\ref{c14_v10_radiography_2} represents a cumulative probability, one\rem{might yield a} \add{can obtain the} probability distribution\rem{function} of imbibition abilities by\rem{simply} calculating the first derivative of $1-f$ with respect to $\Gamma$. For this purpose we used\rem{the} \add{a} polynomial fit \add{to $\Gamma(1{-}f)$}\rem{of the $f$-$\Gamma$-relationship}, shown in Fig.~\ref{c14_v10_radiography_2}. The\rem{resultant} \add{resulting} distribution of $\Gamma$ is shown in Fig.~\ref{c14_v10_radiography_3} for both V5 and V10. 

\rem{However,} One has to keep in mind that the\rem{just} calculated probabilities are volume weighted as opposed to the number weighted probabilites of the pore radii in Fig.~\ref{fig2_3}. This is because \add{the absorption of neutrons is proportional to the volume}\rem{of the volume sensitivity of the regarded signal, that is the adsorbtion of neutrons}. Therefore, a number of {\it large} pores will produce a much higher signal than the same number of {\it smaller} pores. We will come back to this point again in the subsequent discussion of the results. 

\section{Discussion}
Based on our results we\rem{may give} \add{provide} answers to several\rem{interesting} questions concerning flow dynamics in extreme spatial confinement in general as well as liquid invasion into sponge-like pore networks in particular. 

\subsection{Nano-rheology of hydrocarbons in mesoporous silica}
Our\rem{rigorous} \add{quantitative} gravimetrical study on the flow of linear hydrocarbons in mesopores (see Fig.~\ref{conductance})\rem{implies} \add{suggests no significant effect of the all-trans molecule length $\ell$ on the overall flow behaviour} over a wide range of\rem{all-trans molecule lengths} $\ell$ \add{values}\rem{no significant influence on the overall flow behaviour}. Interestingly, we found\rem{hints} \add{indications} for significant deviations from the\rem{arithmetic} \add{constant} average \add{value}\rem{over all alkanes} towards\rem{increased} \add{faster} overall imbibition dynamics only for the largest molecule (n-hexacontane, n-\ce{C60H122}, $\ell \approx 7.6~$nm) in the smallest pores (V5, $r_0 \approx 3.4~$nm). One might\rem{only} speculate\rem{whether} \add{that} the extreme\rem{restrictions} \add{confinement} induce\add{s} \rem{y} alignment or configurational changes \cite{Martin2010a} of the molecule\add{s}\rem{chains}, thereby, enhancing the imbibition ability.\rem{At least,} This result\rem{can be seen} \add{is} in analogous to other imbibition studies that suggest\rem{hints} \add{faster} flow kinetics of polymers in comparable restricted geometries \cite{Shin07, Priezjev2004, Dimitrov2007, Hatzikiriakos2012}.

\rem{An additional} \add{A further} interesting result\rem{are} \add{is} the decreased dynamics of the branched alkane squalane. The imbibition ability is lowered for both V10 (\add{reduction by} 14~\%\rem{ lessening}) and V5 (\add{reduction by} 18~\%\rem{ lessening}). In comparison with the n-alkane of equal length\add{,} that is n-tetracosane (n-\ce{C24H50})\add{,} the additional six side branches (regularly arranged methyl groups) seem to significantly influence the liquid's dynamic behaviour in the pore confinement. This phenomenon can be\rem{well} explained within the framework of\rem{the concept of} sticking layers adjacent to the pore walls that will be part of a more quantitative analysis of the results below.

\rem{Beforehand, though,} We now compare gravimetrical and neutron radiography measurements. From Fig.~\ref{c14_v10_radiography_3} one can\rem{read} \add{extract} the most probable values of the imbibition ability as obtained from\rem{the latter method to} \add{neutron radiography;} $118.5\times 10^{-7}~\sqrt{\rm m}$ for V5 and $179.0\times 10^{-7}~\sqrt{\rm m}$ for V10, respectively. By contrast, the mass increase measurements are not sensitive to a quantity such as {\it the one} most probable imbibition rate but rather to \add{the mass (or volume)-weighted average} $\overline{\Gamma}_V$\rem{that is a mass (or equivalent: volume) weighted average over all existent imbibition abilities}. This\rem{fact elucidates} \add{is due to} the mass' nature as an\rem{integrated} \add{integral} quantity.\rem{At this point} \add{In this respect}, the volume weighted probabilities $P_V$ in Fig.~\ref{c14_v10_radiography_3} are advantageous as they directly allow the calculation of $\overline{\Gamma}_V = {\sum (\Gamma \, P_V(\Gamma)})/{\sum P_V(\Gamma)}$. This\rem{procedure finally} yields $\overline{\Gamma}_V = 116.4\times 10^{-7}~\sqrt{\rm m}$ for V5 and $\overline{\Gamma}_V = 176.0\times 10^{-7}~\sqrt{\rm m}$ for V10,\rem{so} only slight reductions with respect to the peak values \add{are obtained}. 
These\rem{quantities} \add{values} may now be compared to the imbibition abilities obtained from the gravimetrical measurements.\rem{Regarding} \add{Compared to} the\rem{already stated arithmetic} averages over all measured alkanes [V5: $\Gamma=(124.8\pm	2.8)\times 10^{-7}~\sqrt{\rm m}$, V10: $\Gamma=(179.8 \pm	2.3)\times 10^{-7}~\sqrt{\rm m}$] the\rem{above-mentioned quantities} \add{values} of $\overline{\Gamma}_V$ are slightly\rem{larger} \add{smaller} for both \add{matrices}\rem{sample types}. However, as the radiography experiments were only performed with n-tetradecane it might be\rem{the more proper way} \add{appropriate} to\rem{rely on merely} \add{consider} the mass increase measurements of n-\ce{C14H30} \add{only}.\rem{Actually,} Within the error margins the corresponding values $(119.1 \pm	3.5)\times 10^{-7}~\sqrt{\rm m}$ and $(179.3 \pm	5.0)\times 10^{-7}~\sqrt{\rm m}$\add{, respectively,}\rem{do} coincide with the \add{values obtained by neutron radiography}\rem{$\overline{\Gamma}_V$s}. Hence, the mass increase experiments\rem{can be very well described by} \add{are consistent with} the neutron radiography measurements \add{as far as the average imbibition abilities are concerned}\rem{ within the deduced data representation in terms of distributions of imbibition abilities}. 

\subsection{Relation between \add{shape of the} imbibition front\rem{shape} and pore size distribution}
\rem{This is an important proof of the concept of} \add{Our findings provide strong evidence that} not\rem{merely} \add{only} {\it one} but rather a whole\rem{series} \add{set} of (microscopic) imbibition fronts\rem{advancing} \add{advance} in the porous matrix\rem{and obeying} and that these are \add{controlled by} the distribution of imbibition abilities, i.e. rise speeds, discussed above. A possible\rem{explanation} \add{consequence} of this phenomenon can be seen in the remarkable similarity between the $\Gamma$ distributions in Fig.~\ref{c14_v10_radiography_3} and the pore radii distributions in Fig.~\ref{fig2_3}. They\rem{jointly own the} \add{exhibit a similar} shape\rem{being} \add{which is} composed of a most probable value ({\it peak}) with a\rem{more} steep\add{er} decrease\rem{up to} \add{towards} larger values and a rather \add{drawn-out}\rem{sustained} decay\rem{down} to\add{wards} smaller values,\rem{thereby engendering} \add{resulting in} a distinct asymmetry. Now an aforementioned statement comes into play: the imbibition ability is, in principle, direct\add{ly} proportional to the square root of the pore radius \add{(Eq.~\ref{eq:Gamma})}. Hence the similarity between both distributions is due to the\rem{plain} fact that the imbibition ability\rem{can be} \add{is} directly\rem{deduced from} \add{controlled by} the pore radius. 

At first glance, this is astonishing in so far as this result may imply that the porous Vycor glass is made\rem{out} up of many independent and intrinsically monodisperse pore networks with pore dimensions\rem{overall} which satisfy the\rem{known} \add{pore} radius distributions. But the pores are interconnected and thereby one would,\rem{at first glance} \add{naively}, assume a compensation of such effects. However, we recently found a broadening of the imbibition front of water in nanoporous Vycor glass  \cite{Gruener2012}. It could be\rem{documented} \add{shown} that the interface width $w(t)$ increases much faster than observed previously for imbibition front broadening in other porous materials, most prominently paper and sand, namely $w(t)\propto t^\beta$ with here $\beta\approx0.5$. Moreover, the neutron radiography measurements revealed that lateral correlations of the invasion front are short-ranged and independent of time. Thus, the long pores in Vycor indeed seem to allow for simultaneous, but independent imbibition of the individual pores.
 
In fact, Sadjadi and Rieger\rem{documented} \add{showed} by simulations of pore networks mimicking the structure of Vycor that spontaneous imbibition crucially depends on the pore aspect ratio \cite{Lenormand1990, Sadjadi2013, Lee2014}. For short pores, neighbouring menisci coalesce and form a continuous imbibition front. Thus the smoothening effect of an effective surface tension within the interface leads to a slow broadening of the front. By contrast, for elongated pores (like in Vycor) individual menisci form\rem{ in pore space}, which results in effectively independent capillary filling events in a three dimensional network of cylindrical pores. See Fig.~\ref{fig:CRBroadening}(a) for an illustration of this meniscus splitting process at a pore junction and the subsequent filling of two elongated pores (capillaries) with small and large diameter. The speed of these filling events is given by the imbibition ability and thus depends on the pore radii, $\Gamma=\Gamma(r)$ according to Eq. (\ref{eq:Gamma}).

The\rem{advancing} light scattering \add{of the advancing} front may also be interpreted by a distribution of imbibition abilities. The visual observation of the\rem{dynamics of this} front \add{relies} on\rem{structures composed} \add{the presence} of a partly filled network of pores. Considering rapidly filling large and\rem{tardily} \add{slowly} filling small pores, the emergence of this phenomenon \add{follows directly from}\rem{is nothing more than a logical consequence of} the considerations above. Comparing the \add{$\Gamma$ values at the} upper and lower bound\add{s}\rem{$\Gamma$ values} of the n-tetradecane front in V5 and V10 (see Tab.~\ref{gammaheight}) with the $f$-$\Gamma$-relationships obtained from the \add{neutron} radiography measurements (see Fig.~\ref{c14_v10_radiography_2}), the emergence of the light scattering phenomenon can be linked to the filling degree $f$. For both\add{,} V5 and V10\add{,} the lower and upper bound are \add{found to be} located at $f\approx (45\pm 5)~\%$ and $f\approx (100)~\%$, respectively, which implies that for \add{$f<45~\%$}\rem{lower filling degrees} too \add{few}\rem{less} pores are filled \add{leading to fluctuations in}\rem{engendering characteristic variation length scales of} the refractive index\rem{higher} \add{on a length scale larger} than the wavelength of visible light and, thus, inducing no light scattering.\rem{Crossing the} \add{Beyond} $f\approx 45~\%$\rem{bound structures} \add{fluctuations} on adequate length scales are\rem{gradually} \add{increasingly} created.\rem{A remarkable fact} \add{It} is \add{remarkable} that such\rem{appropriate structures are present} \add{fluctuations exist} up to high filling degrees as indicated by the lower bound's imbibition abilities. 



\subsection{Determination of pore size and hydrodynamic radii distribution\add{s} from imbibition front broadening} 
Finally, we\rem{may} turn to a more quantitative interpretation of the obtained results. Applying Eq.~\eqref{eq:Gamma} we \add{can} gain access to the hydrodynamic pore radii $r_{\rm h}$\rem{given that} \add{if} we can\rem{assess} \add{estimate} the Laplace radius $r_{\rm L}$.\rem{Having a close look at} Eq.~\eqref{eq:Gamma}\rem{one might estimate its overall impact} suggests that $r_{\rm L}$ has a smaller effect on the dynamics than $r_0$ and particularly $r_{\rm h}$. But, a priori, there are no hints on\rem{its} \add{the} exact value \add{of $r_{\rm L}$}.

\rem{With regard to} \add{Based on} water sorption isotherms in porous Vycor (see Fig.~\ref{water_isos}) that\rem{express} \add{reveal} the thermodynamics of water in the mesopore confinement\rem{ in detail}, we may\rem{only} conclude that under standard laboratory conditions ($10~\% \lesssim {\rm RH} \lesssim 40~\% i.e. \; 0.1 \lesssim p \lesssim 0.4 $) the sample's maximum filling degree is 25~\% corresponding to approximately 2 layers of water molecules. According to these considerations $r_{\rm L}$ should be\rem{reduced by maximal} \add{at most} 0.5~nm (roughly two times the diameter of a water molecule)\rem{with respect to} \add{smaller than} $r_0$.
\begin{figure} \center
\includegraphics[width=.8\columnwidth]{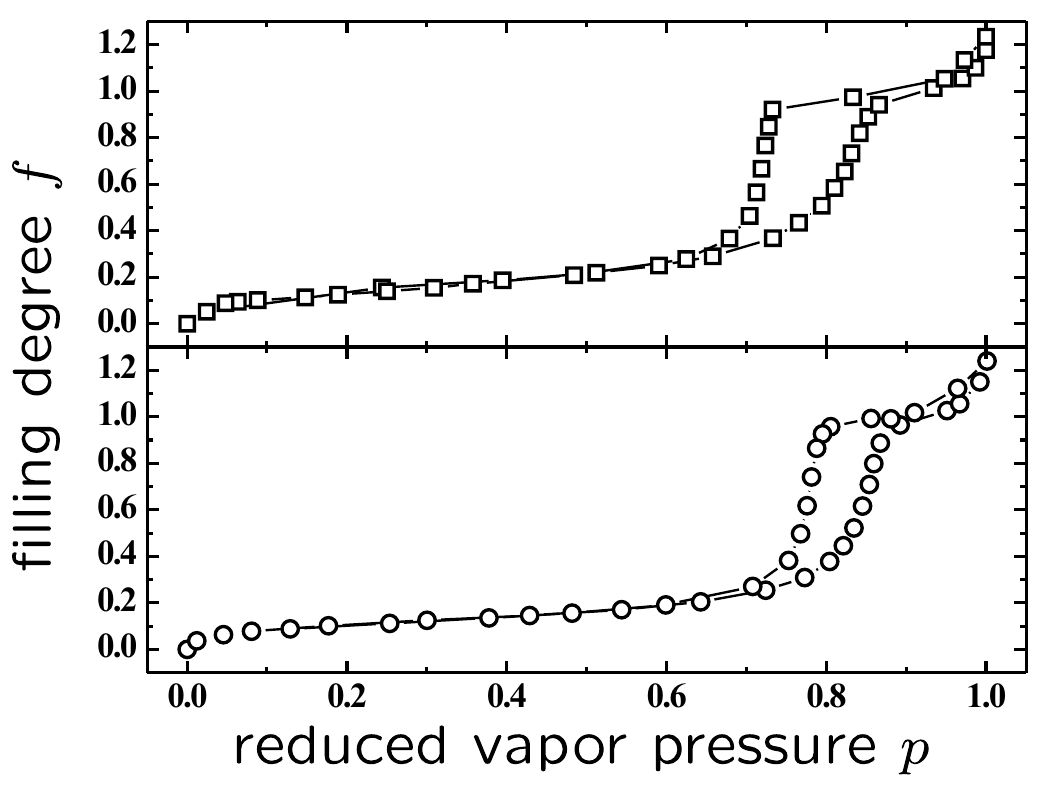}
\caption{Water sorption isotherms of V5 (upper panel) and V10 (lower panel) recorded at $4~^\circ$C. The reduced vapor pressure $p$ is identical to the relative humidity (RH). The data points with $f>1$ indicate the formation of bulk water droplets outside the mesopores.}
\label{water_isos}
\end{figure} 

In order to estimate the\rem{actual influence} \add{effect} of $r_{\rm L}$ on the overall dynamics we refer to the imbibition measurements of n-hexa\add{decane}\rem{contane} (n-\ce{C16H34}) imbibing into V5 as a function of the humidity\rem{in the laboratory} as discussed in section IV A - see Tab.~\ref{gammaC16} and Tab.~\ref{gammaheight}.\rem{It is obvious that even} Relative humidities of up to 50~\% do not\rem{distinctly} \add{significantly} influence the dynamics of the mass increase or (roughly equivalent) the upper bound of the light scattering front\add{,} provided one takes into account the gradual and significant decrease in the initial porosity $\phi_{\rm i}$ with increasing humidity. Interestingly, the lower bound of the light scattering front noticeably slows down with\rem{rising} \add{increasing} humidity. This is\rem{not too surprising as the impact of} \add{due to the fact that} a coverage with a certain thickness affects smaller pores far more than larger pores. \add{Thus,} whereas\rem{this effect seems not to influence} the\rem{many} \add{numerous} larger pores\rem{at all} \add{are hardly affected}, it has a\rem{distinct} \add{significant} impact on the small pores that show\rem{decreased} \add{reduced} dynamics so that the light scattering front extends to lower values. However, as the small pores are only few and carry only little fluid\add{,}\rem{at all} this phenomenon does not influence the overall dynamics of the rise process. We may, therefore, assume that the \add{magnitude of } $r_L$ only plays a\rem{subordinate} \add{minor} role. Nonetheless, for the\rem{subsequent} \add{following} analysis we need to quantify $r_{\rm L}$.\rem{Referring to the insights deduced from} \add{Based on} the water sorption isotherms we fix the Laplace radius to $r_{\rm L} = r_0 - 0.25$~nm,\rem{a value reduced} \add{corresponding to a reduction caused} by approximately one monolayer of water at the pore walls.

Now, we can turn to the relation of the distribution functions of $\Gamma$ and $r_{\rm h}$. The direct connection between the pore radius $r_0$\label{radius2} and the imbibition ability $\Gamma$ via Eq. \ref{eq:Gamma} allows for a more fundamental analysis of the relation between their distribution functions. With $r_0 \to r_h+\Delta$ and $r_L \to r_0-0.25$\,nm\ one obtains
\begin{equation}
\Gamma(r_h) = \frac{r_h^2}{r_h+\Delta}\,\sqrt{\frac{\phi_0}{\tau\,(r_h+\Delta-0.25\,\text{nm})}}.
\label{eq:GammaRadiusRelation}
\end{equation}
Hence, the $P(\Gamma)$-distribution can be transformed in a $P(r_h)$-distribution with $P(\Gamma)\, d\Gamma=P(r_h)\,d{r_h}$, see Fig.~\ref{fig2_3} for the resulting $P(r_h)$-distribution. From this plot, it is evident that the \add{distribution of} hydrodynamic radii determined from the imbibition front broadening is in remarkable quantitative agreement with the pore size distributions derived from the nitrogen sorption isotherm measurements, provided one allows for a fixed shift $\Delta$ with $\Delta=0.63~$nm and $\Delta=0.52~$nm for V5 and V10, respectively.

Thus, the imbibition front broadening can indeed be described by independent filling events of cylindrical pores with a size distribution typical of the Vycor glasses investigated. Unfortunately, we performed a detailed pore size analysis based on imbibition front broadening for C14 only and we do not have corresponding data sets for the other hydrocarbons. However, since our analysis is not specific to C14, we believe that the imbibition front analysis presented here can also be applied to other hydrocarbons (or liquids).

We would like to stress that in a network of connected elongated pores (or capillaries) with random radii another important mechanism has to be considered. As Sadjadi and Rieger showed, at pore junctions, where menisci with different Young-Laplace and thus imbibing pressures compete\add{,} meniscus propagation in one or more branches (depending on the pore size distribution) can come to a halt, when the negative Young-Laplace pressure of one of the menisci exceeds the hydrostatic pressure within the junction \cite{Sadjadi2013, Lee2014, Sadjadi2015, Rieger2015}. These\rem{menisci} arrests \add{of the menisci} persist until the hydrostatic pressure in the junction is larger than the Young-Laplace pressure of the\rem{halted} \add{arrested} menisci, see Fig. \ref{fig:CRBroadening}(b).\rem{Caused by} \add{Due to} the viscous drag in the liquid column behind the advancing menisci, the junction pressure increases linearly with the distance of the moving menisci from the junction and this distance follows the Lucas-Washburn dynamics. Hence, the arrested menisci start to move and thus to fill the corresponding pore (or pore segment), when the moving menisci have advanced by a distance, which is proportional to $\sqrt{t}$ \cite{Sadjadi2013, Lee2014, Sadjadi2015, Rieger2015}. Note that they move then with a velocity larger than the menisci in the small pores, because of their larger $\Gamma$s. 

Unfortunately, there is no thorough analysis of the meniscus arrest time statistics and its influence on the front shape evolution as a function of pore size distribution available up to now. However, the menisci arrests result solely in delays of the filling events with delay times that increase with the pore size and with time. They do not prevent the eventual occurrence of those filling events, except for later stages of the filling process \cite{Sadjadi2013, Lee2014}, when negligible amounts of liquids have to be transported. Then capillarity completely determines the liquid menisci advancement and final position in pore space (capillarity-determined imbibition front broadening). By contrast, the sizeable liquid flow in the porous medium at earlier imbibition times means that both hydraulic permeability and capillarity, quantified by the imbibition ability, determine the liquid supply from the bulk reservoir and menisci advancement, and hence the imbibition front broadening. It is conceivable that a transition between this imbibition ability-determined and a capillarity-determined broadening regime exists. This would also mean that initially preferentially percolating liquid-filled clusters of interconnected larger pores form (with low hydrostatic pressure drops for a given flow rate), whereas later the imbibition front is determined by the occurrence of arrested large menisci and thus empty segments of large pores. Such a transition would be analogous to the situation encountered for imbibition in a set of parallel, independent capillaries in a gravitational field, where initially the menisci in the large pores propagate the fastest and are ahead of the ones in the small capillaries. Whereas finally, upon reaching the final capillary rise heights the viscous drag and hence hydraulic permeability is negligible and a balance of capillarity and gravitation implies an opposite arrangement of the menisci as a function of height, known as Jurin's law \cite{Gennes2004}, i.e. the smaller the pore the higher the liquid will rise, $h \sim 1/r_0$. 

Our analysis suggests that in the present case the broadening is determined by the imbibition ability, and thus by both a competition of hydraulic permeability and capillarity. Since this phenomenology as well as menisci arrests result in a $\sqrt{t}$-broadening of the front, they are difficult to distinguish. Additional studies are clearly necessary to explore the interplay and competition of imbibition ability- and meniscus arrest-determined broadening and its possible dependence on the pore morphology, in particular the aspect ratio of the pores. This is also important in order to decide whether the evidence of the validity of our model is rather a coincidence or rests on the proper mechanistic insights with regard to the pore filling events in Vycor. 


\subsection{Partitioning in interface/core fluidity}
Another important insight of Fig. \ref{fig2_3} is the shift between the $r_h$ and the $r_0$ distribution. It implies that a liquid layer of thickness $\Delta$ adjacent to the pore walls does not participate in the flow. Rather, it sticks to the wall or shows\rem{tremendously decreased} \add{a significantly reduced} mobility whilst the\rem{remaining contingent's} dynamics \add{of the remaining liquid} obey\add{s}\rem{the law of} Hagen-Poiseuille\add{'s law} and, therefore, the classical assumptions of continuum mechanical theory. This reveals that the imbibed liquid is separated in two regions (1) an interfacial layer (with thicknesses $\Delta=0.63~$nm and $\Delta=0.52~$nm for V5 and V10, resp\add{ectively}) whose \add{arrested or very slow} kinetics are mainly determined by the interaction between liquid and substrate and (2)\rem{an inner (away from the interface)} \add{a central} region that\rem{shows the} \add{follows} classical behaviour as\rem{predicted from collective} \add{expected for bulk} liquids\rem{properties}. All these macroscopic concept\add{s remain}\rem{stay} valid for this\rem{inner compartment} \add{central region} even though it comprises merely 100 to 1000 molecules per cross-sectional area,\rem{so being} far fewer than \rem{below the concregations of $10^{23}$ molecules} typical\rem{ly preconditioned} \add{ensembles considered} in statistical physics}. 

The behaviour of the interfacial layer is\rem{deeply related to} \add{dominated by} the liquid/substrate interaction. In this context one might recall that we have already considered an initial water coating of the pore walls due to the\rem{final} \add{finite} humidity. This effect\rem{preeminently proving} is \add{caused by} Vycor's extremely high hydrophilicity so that the\rem{obtained result} \add{existence} of a sticking layer is not too surprising. Especially the first adsorbed water layer is stabilized by the attractive potential between silica and water. This is also\rem{expressed by} \add{reflected in} the distinct monolayer step in the sorption isotherms (Fig.~\ref{water_isos}) and a slow self-diffusion dynamics \cite{BellissentFunel1993, Chen1995}. Molecular Dynamics studies also document a glassy structure of the water boundary layer in Vycor \cite{Gallo2002} and the existence of sticky boundary layers has been inferred from Hagen-Poiseuille nanochannel flows  \cite{Heinbuch1989}. Moreover, a water coating on the pore walls is far more \add{stable}\rem{stabilized} than an alkane coating (as\rem{expressed} \add{indicated} by their surface tensions), so that the imbibed hydrocarbon is not able to displace\rem{this} \add{the} initial water \add{layer}\rem{coverage}. Hence, we\rem{have to assume} \add{suggest that} this monolayer\rem{to be} \add{is} present during the whole experiment. According to previous publications \cite{BellisentFunel1993, Chen1995, Gruener2009} it is extremely immobile\add{,} thereby providing a first \add{layer}\rem{contribution} of \add{thickness} $\sim 0.25$~nm to the overall sticking layer thickness $\Delta$. 

However, this water layer alone\rem{might} \add{can} not \add{constitute}\rem{explain} the complete sticking layer\rem{ thickness}. The remaining 0.3~nm to 0.4~nm must originate\rem{in} \add{from} an additional immobile layer. This thickness\rem{saliently} coincides with the diameter of a hydrocarbon chain in all-trans configuration that is $\sim 0.35$~nm. Hence, we\rem{may} assume \add{that} the sticking layer\rem{to be constituted} \add{consists} of a water layer directly adjacent to the pore walls\rem{and an ensuing} \add{covered by a} layer of flat\add{-}lying hydrocarbon chains. This immobile shell is in agreement with the pioneering experiments \add{by Debye and Cleland} on forced imbibition of n-alkanes\rem{by Debye and Cleland} \cite{Debye1959}, consistent with studies regarding the thinning of n-alkane films in the surface force apparatus \cite{Chan1985, Christenson1982, Stevens1997, Georges1993, Heinbuch1989} and beam-bending experiments on the rheology of alkanes in Vycor \cite{Scherer2000, Vichit-Vadakan2000}. Moreover, X-ray reflectivity studies have indicated one strongly adsorbed, flat\add{-}lying monolayer of hydrocarbons on silica \cite{Basu2007, Mo2005, Bai2007, Corrales2014}. Quasi-elastic neutron scattering measurements, which are sensitive to the center-of-mass self-diffusion of the n-alkanes in the pores and thus the liquid's viscosity, also indicate a partitioning of the diffusion dynamics of the molecules in the pores in two species: One component with\rem{a} bulk-like self-diffusion dynamics and a second one which is immobile, \add{and thus `}sticky\add{',} on the time scale probed in the neutron scattering experiment \cite{Baumert2002, Kusmin2010, Kusmin2010a, Hofmann2012}. 

This concept is further corroborated by the\rem{already presented} squalane measurements. The $\sim 15~\%$ decrease in the imbibition ability and the\rem{consequent} \add{corresponding} additional reduction of the hydrodynamic pore radii by approximately $0.3$~nm can be\rem{consistently} explained by the larger diameter of the squalane molecule as compared \add{to}\rem{with} a non-branched hydrocarbon chain. Its extra methyl groups\add{,} regularly and alternately attached to the n-tetracosane backbone\add{,} cause an increase of the sticking layer's thickness by a value comparable to the\rem{just} \add{above-}mentioned decrease in $r_{\rm h}$. 

\section{Conclusions}
We presented experiments on the capillarity-driven invasion dynamics of hydrocarbons in mesoporous, monolithic silica glass. The invasion kinetics for all liquids investigated are governed by classical Lucas-Washburn laws typical of spontaneous imbibition in porous media. The pre-factors of the corresponding square-root-of-time behaviours are compatible with the bulk fluid parameters, provided we assume \add{a}\rem{one sticky} \add{sticking} boundary layer\rem{corresponding to the thickness of a} of flat-lying hydrocarbon\add{s}\rem{backbone} and a monolayer of water adsorbed at the pore walls. For the molecule with \add{the} longest chain investigated, C60\add{,} an increased fluidity in the mesopores is suggested by the results. This may indicate the transition towards polymer-like flow behavior\rem{ in the pores} \cite{Dimitrov2007, Mueller2008}, i.e. the absence of a pore-wall adsorbed molecular layer in combination with velocity slippage, shear-induced alignment or rearrangements in the molecular structure which result in an increased mobility compared to the bulk state. These rheological insights are in good agreement with previous experiments on water, alkanes and liquid crystals \cite{Huber2007, Gruener2009, Gruener2009a, Gruener2011, Vincent2015}. They are also corroborated by Molecular Dynamics simulations on spontaneous imbibition in nanopores \cite{Gelb2002, Dimitrov2007, Chibbaro2008}. 

\rem{Moreover, pronounced light scattering at the advancing imbibition front indicates a substantial broadening of the invasion front for all liquids studied.} Both, light and neutron imaging reveal not only filling kinetics which follow a square-root-of-time dependence, \add{but} also that the overall width of the front obeys this\rem{kinetics} \add{time-dependence}, as previously reported for water invasion in Vycor \cite{Gruener2012}.\rem{In fact,} Neutron radiography allowed us to \add{temporally and} spatially resolve the evolution of the asymmetric shape of the imbibition front. The imbibition front shape can be described by a superposition of independent wetting fronts moving with radius-dependent square-root-of-time laws, where the contribution of the single fronts is weighted by a distribution of \textit{hydrodynamic} pore diameters that agrees with the pore size distribution obtained from volumetric nitrogen sorption isotherm measurements, when corrected for the immobile\rem{, sticky} boundary layer\rem{ thickness}. \add{The superposition of these independent fronts leads to fluctuations on the length scale of visible light and results in the observed light scattering.} 

Our finding\add{s} suggest\rem{s} that the analysis of imbibition fronts allows independent determination of pore size distributions\rem{ for this porous medium}, provided one corrects the distributions for a possible immobile boundary layer, which does not participate in the viscous flow. The\rem{origins of} \add{reason for} this relation\rem{are} \add{is} traced \add{back} to the\rem{very} \add{same} mechanisms responsible for imbibition front broadening, in particular independent meniscus movements and independent capillary filling events as typical for spontaneous imbibition in networks of elongated pores with random radii  \cite{Gruener2012, Sadjadi2013, Lee2014, Sadjadi2015, Rieger2015}. We believe that further model calculations and simulation studies on artificial pore networks along with a direct experimental study of pore-scale air displacement statistics (menisci arrests and menisci advancement in single pores), e.g. by time-dependent, in-situ neutron \cite{Gruener2012} or X-ray imaging\cite{Berg2013, Murison2014}  or small-angle scattering, are necessary to clarify the relation between imbibition front broadening, the pore size distribution, and the aspect ratio of the pores.

Many porous media have a complex geometry (variation of the mean pore diameter within isolated channels, meandering of the pores, varying pore connectivity) \cite{Sahimi1993, Halpinhealy1995, Hinrichsen2000, Alava2004}. Similarly, as observed here, the inhomogeneities result in variations in the local bulk hydraulic permeability and in the capillary pressure at the moving interface resulting in \add{a} broadening of the imbibition front and thus the coexistence of empty and filled pore segments during the imbibition process. Often this broadening displays universal scaling features on large length and time scales, which are independent of the microscopic details of the fluid and the matrix \cite{Kardar1998, Spathis2013, Planet2007, Dube2007, Buldyrev1992, Horvath1995, Miranda2010, Hernandez-Machado2001, Geromichalos2002, Leoni2011}. In the past\add{,} most imbibition front broadening studies focused on sand, packed beads and paper. In these systems, the pores are not elongated, \add{and hence} no individual menisci form in neighbouring pores. This results in a continuous liquid-gas interface, whose advancement is spatially correlated due to an effective surface tension \cite{Dube2000}. Consequently\add{,} menisci advancement beyond the average front position is slowed down whilst menisci lagging behind are drawn forward. Thus, in contrast to the situation present\add{ed}\rem{s} here, the front shape does not contain any\rem{details of} \add{information on} the pore size distribution. 

The determination of pore sizes and pore morphologies is arguably one of the most important scientific and technological topics in the field of porous media \cite{Rouquerol1998}. Depending on the type of porous material and the experimental conditions\rem{ met}, such diverse techniques as direct imaging \cite{Rouquerol1998, Gommes2009}, neutron and x-ray scattering \cite{Hofmann2005, Zickler2006, Mascotto2009, Bon2014}, gas sorption \cite{Naumov2009, Morishige2003, Cychosz2012, Landers2013, Thommes2014}, monitoring of phase transition shifts (cryporometry) \cite{Brun1977, Huber1999, Landry2005, Riikonen2011}, measuring of imbibition kinetics \cite{Elizalde2014}, gas pressure-induced displacement of wetting liquids \cite{Broncano06} can be appropriate means to achieve this goal. Most of these methods rely on size-effects in the thermodynamics, expressed by the radius-dependent Kelvin equation (capillary condensation) or Gibbs-Thomson equation (cryoporometry) and thus on equilibrium \add{properties}\rem{ physics}. Interestingly, the experiments presented here suggest that also a non-equilibrium physical phenomenon can contribute to the field of porosimetry. In the case of individual moving liquid menisci, as in the case of elongated pores where menisci are split at pore junctions, the broadening of spontaneous imbibition fronts of wetting liquids constitutes an alternative way for the determination of pore size distributions. To what extent heterogeneities in the wettability of the porous medium\cite{Murison2014}, as is often the case in porous rocks, affect this phenomenology is an interesting topic to explore in the future. We also hope that our study stimulates a detailed exploration of the conditions for a reliable applicability of this approach by experiments on real \cite{Berg2013} and model pore systems \cite{Courbin2007,Datta2014a, Sadjadi2015, Rieger2015}, but also by computer simulations and phenomenological model calculations on spontaneous imbibition in pore networks \cite{Lam2000, Sadjadi2013}.

This work has been supported within the DFG priority program 1164, \textit{Nano- \& Microfluidics} (Grant. No. Hu 850/2). We thank Zeinab Sadjadi and Heiko Rieger for stimulating discussions and the research neutron source Heinz Maier-Leibnitz (FRM II) in Munich (Germany) for beam time at the imaging experiment ANTARES. 
\bibliographystyle{unsrtnat}

\end{document}